%% file: acl_latex.tex
\pdfoutput=1
\documentclass[11pt]{article}

\usepackage[preprint]{acl}

\usepackage{times}
\usepackage{latexsym}
\usepackage[T1]{fontenc}
\usepackage[utf8]{inputenc}
\usepackage{microtype}
\usepackage{inconsolata}
\usepackage{graphicx}
\usepackage{xcolor}
\usepackage{listings}
\usepackage[most]{tcolorbox}
\usepackage{tikz}
\usetikzlibrary{arrows.meta,positioning,calc,shapes.geometric,shadows.blur,fit,backgrounds}
\usepackage{float}
\usepackage{placeins}
\usepackage{multirow}
\usepackage{dblfloatfix}
\usepackage[colaction]{multicol}
\usepackage{afterpage}

\newcommand{\thickhline}{\noalign{\hrule height 1.1pt}}

\usepackage{xspace}
\newcommand{\Name}{Evoke-Sim\xspace}
\title{Evaluation of Motivational Interviewing Counsellors with Task-Aware Multi-Stage LLM-Based Simulated Clients}

\author{
  \textbf{Jiading Zhu\textsuperscript{*}},
  \textbf{Xinyu Cindy Wang\textsuperscript{*}},
  \textbf{Thomas Nguyen\textsuperscript{*}},
  \textbf{Yan Qing Lee\textsuperscript{*}}
  \\
  \textbf{Osnat C. Melamed\textsuperscript{*\textdagger}},
  \textbf{Peter Selby\textsuperscript{*\textdagger}},
  \textbf{Jonathan Rose\textsuperscript{*\textdagger\textsection}}
  \\
  \textsuperscript{*}University of Toronto
  \\
  \textsuperscript{\textdagger}Centre for Addiction and Mental Health,
  Toronto, ON, Canada
}

\begin{document}
\maketitle
\begingroup
\renewcommand{\thefootnote}{\fnsymbol{footnote}}
\footnotetext[4]{Corresponding author:
  \href{mailto:jonathan.rose@utoronto.ca}{jonathan.rose@utoronto.ca}}
\endgroup
\begin{abstract}
The development and benchmarking of Large Language Model (LLM)-based Motivational Interviewing (MI) counsellors now often rely on LLM-based simulated clients. Prior work on simulated clients, however, has not aligned with the specific \emph{tasks} fundamental to the MI therapy approach. A key task is \emph{evoking}, in which the counsellor first elicits the client's ambivalence and then strengthens the client's motivation for change. We present \textbf{\Name}, a task-aware, multi-stage LLM-based client simulation framework for evaluating MI counsellors in smoking cessation, designed specifically for the evoking MI task. \Name employs structured client profiles, an evoking-specific three-stage conversation flow, and a \emph{reveal} policy that regulates which client profile information might be disclosed at each stage. We show that compared to existing profile-grounded simulated clients, \Name is better at differentiating levels of MI quality using task-aware evaluation metrics, while reducing non-grounded client statements and premature disclosure of client information, setting a higher standard for the evaluation of LLM-based MI counsellors. 
\end{abstract}

\input{introduction}
\input{related_works}
\input{methodology}
\input{experiments_results_discussion}
\input{conclusions}
\input{limitations}
\input{ethical_considerations}

\bibliography{custom}

\clearpage
\appendix
\nolinenumbers
\onecolumn
\newcommand{\setappendixnumbering}[1]{%
  \setcounter{table}{0}%
  \setcounter{figure}{0}%
  \setcounter{equation}{0}%
  \setcounter{lstlisting}{0}%
  \renewcommand{\thetable}{#1.\arabic{table}}%
  \renewcommand{\thefigure}{#1.\arabic{figure}}%
  \renewcommand{\theequation}{#1.\arabic{equation}}%
  \renewcommand{\thelstlisting}{#1.\arabic{lstlisting}}%
}

\clearpage
\onecolumn
\nolinenumbers
\setappendixnumbering{A}
\input{profile_extraction_multicolumn}
\clearpage
\onecolumn
\nolinenumbers
\setappendixnumbering{B}
\input{change_score_multicolumn}
\clearpage
\onecolumn
\nolinenumbers
\setappendixnumbering{C}
\input{architecture_modules_multicolumn}
\clearpage

\onecolumn
\nolinenumbers
\setappendixnumbering{D}
\input{classifier_prompts_multicolumn}
\clearpage

\onecolumn
\nolinenumbers
\setappendixnumbering{E}
\input{state_control_multicolumn}
\clearpage

\onecolumn
\nolinenumbers
\setappendixnumbering{F}
\input{reveal_control_multicolumn}
\clearpage

\onecolumn
\nolinenumbers
\setappendixnumbering{G}
\input{prompt_builder_prompts_multicolumn}
\clearpage

\onecolumn
\nolinenumbers
\setappendixnumbering{H}
\input{profile_only_client_multicolumn}
\clearpage

\onecolumn
\nolinenumbers
\setappendixnumbering{I}
\input{mi_nonadherent_counsellor_multicolumn}
\clearpage

\onecolumn
\nolinenumbers
\setappendixnumbering{J}
\input{mibot_v63a_counsellor_multicolumn}
\clearpage

\onecolumn
\nolinenumbers
\setappendixnumbering{K}
\input{mibot_evoke_counsellor_multicolumn}
\clearpage

\onecolumn
\nolinenumbers
\setappendixnumbering{L}
\input{grounding_prompts_multicolumn}

\clearpage

\onecolumn
\nolinenumbers
\setappendixnumbering{M}
\input{omnibus_tests_multicolumn}

\clearpage

\onecolumn
\nolinenumbers
\setappendixnumbering{N}
\input{grounding_all_settings_multicolumn}

\clearpage

\onecolumn
\nolinenumbers
\setappendixnumbering{O}
\input{rq2_pairwise_tests_multicolumn}

\end{document}

%% file: introduction.tex
\section{Introduction}

The application of Large Language Models (LLMs) for automated counselling has become an active research direction, especially for mental health support and psychotherapy \citep{info:doi/10.2196/57400, huascopingreview, na-etal-2025-survey}. A key challenge in this work is to properly evaluate counselling quality against established methodologies and standards. This is now urgent as the wide availability of general-purpose chatbots has led to their widespread use by the general public for mental health counselling \citep{rousmaniere2025large}. The focus of this paper is the evaluation of behaviour change counselling using Motivational Interviewing (MI) \citep{MillerRollnick2023}, through the use of simulated clients.

MI is widely-used by human counsellors in domains such as smoking cessation and alcohol reduction \citep{Lindson10, NYAMATHI201023}. MI counsellors work through four tasks when speaking with a client: "Engaging", "Focusing", "Evoking", and "Planning". A key task is "Evoking", where the counsellor elicits and acknowledges the client's ambivalence towards changing the unhealthy behaviour and gently guides the client towards change by eliciting the client's own motivations for change.

Several recent studies have already explored and achieved strong results in instructing LLMs to role-play as MI counsellors \citep{xie-etal-2024-shot-dialogue,sun-etal-2025-rethinking,mahmood-etal-2025-fully,yang-etal-2025-cami}. While the performance of these LLM-based MI counsellors can be naively evaluated by directly comparing generated counsellor utterances against human transcripts, proper evaluation requires interactions with real clients that are ambivalent about change. Although recruiting real clients to participate in human studies is possible \citep{andrewMI, bickmoreMI, mahmood-etal-2025-fully}, the recruitment process and the experiments themselves are often long and costly, and raise practical and ethical concerns when participants are exposed to unqualified LLM-counsellors. 

One solution is to rely on LLM-based simulated clients, as explored in recent works \citep{yosef-etal-2024-assessing,wang-etal-2024-patient, chiu2024computationalframeworkbehavioralassessment, wang2024clientcenteredassessmentllmtherapists, kim-etal-2025-kmi,yang-etal-2025-consistent}. Most of these approaches are grounded on conversations with human participants, either by reusing counselling transcripts as in-context examples, or by extracting structured client profiles from the participants. These synthetic clients can then engage in conversations with LLM-based counsellors, generating transcripts that must be rigorously evaluated. Common ways of evaluation include behavioral coding tools such as MISC \citep{MISC}, MITI \citep{MITI} and CLEAR \citep{CLEAR}, for which recent work has reported success in automation with the help of language models \citep{MI-TAGS, BIMISC, AutoMISC}.

Despite these promising developments, automated evaluation of MI counselling quality remains a challenge: even though behavioral coding can be automated, session-level evaluation depends on summary scores such as "percentage of complex reflections made" or "reflection-to-question ratio" \citep{MITI}, which only captures part of the picture. 
The appropriate choices of counsellor strategies depend on the client and the context, as well as the specific MI task at hand. Therefore, to properly evaluate MI counselling quality, it is crucial to assess not only the presence of MI-adherent skills, but also their timing and content in relation to the conversation and the MI task.

To address this challenge, we present \textbf{\Name}, a novel LLM-based client simulation framework designed for task-aware, multi-stage evaluation of MI counsellors. We have focused on the evoking task as a first step because it is central to behavior change and the prerequisite for moving to planning. We partition the evoking task into three connected stages in consultation with MI experts. \Name grounds each simulated client in a structured profile parsed from counselling sessions with human clients, and uses a multi-stage conversation flow to model progression within the evoking task. Most importantly, \Name implements stage-dependent client instructions and information disclosure mechanisms that constrain what the client can disclose and when, enabling task-aware evaluation of timing and content of counsellor behaviours.

Our main contributions are:
\begin{itemize}
  \item \Name, a profile-grounded client simulation framework for task-aware, multi-stage evaluation of MI counsellors on the evoking MI task (code released).
  \item A set of task-aware, stage-specific evaluation metrics across the three stages of evoking.
  \item We show that \Name better differentiates MI counselling quality than generic profile-based simulated clients and prior metrics.
  \item We demonstrate that \Name reduces non-grounded statements and premature disclosure relative to generic profile-based simulated clients.
  \item A released dataset of 83 diverse and structured profiles extracted from MI counselling sessions with human participants.
\end{itemize}

%% file: related_works.tex
\section{Related Work}
\label{sec:related_works}

Recent work has explored using LLM-based simulated counselling clients to evaluate LLM-based MI counsellors, with various approaches to ground the conversation. BOLT \citep{chiu2024computationalframeworkbehavioralassessment} grounds their simulated clients with real MI transcripts and evaluates LLM-based MI therapists with behavior analysis. Both \citet{yosef-etal-2024-assessing} and ClientCAST \citep{wang2024clientcenteredassessmentllmtherapists} ground their simulated clients with structured profiles, either with a variation of pre-defined parameter settings or extracted from transcripts of real counselling sessions, and evaluate the quality of simulated sessions using LLM-filled questionnaires. \citet{kim-etal-2025-kmi} ground their simulated clients through in-context learning with web-crawled context data, and specifically instructs the client to output change talk. 

Most recently, structured client internal states have been shown to improve the consistency of LLM-based simulated clients; \citet{yang-etal-2025-consistent} proposed a consistent MI client simulation framework that, in addition to grounding with client profiles, defines client internal states based on the trans-theoretical model of health behavior change \citep{TTM}, and explicitly tracks client state transitions to keep generated client utterances aligned with profile constraints throughout entire sessions. Beyond MI, PATIENT-$\psi$ \citep{wang-etal-2024-patient} conditions simulated patients on explicit cognitive models from cognitive behavior therapy (CBT) \citep{CBT}, supporting that structured internal-state grounding can improve the realism of simulated patients.

\Name builds onto existing methodology for simulated clients and provides unique and novel contributions: we extend profile-based grounding with a rule-based information disclosure policy to control what client information can be revealed at each turn, and model client internal states as a multi-stage conversation flow aligned with the evoking MI task.

%% file: methodology.tex
\section{Methodology}
\label{sec:methodology}

This section describes the design of \Name, a task-aware, multi-stage framework for simulating clients and evaluating MI counsellors within the "Evoking" task of MI.

\Name has three key design contributions:
\begin{enumerate}
    \item \textbf{Structured profiles.} Each simulated client is anchored to a structured profile extracted from human counselling data, so responses are tied to realistic persona, motivations, barriers, and potential next steps.
    \item \textbf{Task-aware multi-stage progression.} Client behavior is controlled through a deterministic three-stage flow with explicit transition and early-termination rules aligned with the evoking MI task.
    \item \textbf{Rule-based reveal policy.} Information disclosure from the simulated client is gated turn by turn, so that only appropriate profile items are revealed.
\end{enumerate}
The following sections give the details of these elements of \Name.
\subsection{Structured Profiles}
\label{subsec:structured_profiles}

\begin{table*}[!t]
\centering
\scriptsize
\setlength{\tabcolsep}{3pt}
\renewcommand{\arraystretch}{1.2}
\begin{tabular}{p{0.17\textwidth}p{0.21\textwidth}p{0.26\textwidth}p{0.24\textwidth}}
\thickhline
\textbf{Stage} & \textbf{Goal} & \textbf{Transition to Next Stage} & \textbf{Early Termination} \\
\hline
Stage 1 \newline(Evoking Ambivalence) & Evoke both sides of ambivalence & At least one motivation and one barrier item evoked from the client & If transition requirements are not met within 10 client turns; if next steps or commitment were asked 3 times\\
\hline
Stage 2 \newline(Evoking Change Talk) & Evoke change talk and reduce sustain talk & Change score $\geq 3$ & If change score reaches the floor; if next steps or commitment were asked 3 times\\
\hline
Stage 3 \newline(Evoking Commitment to Next Steps) & Evoke potential next steps, then commitment. & Commitment+ is evoked after any potential next steps have been discussed (end goal) & If change score reaches the floor \\
\thickhline
\end{tabular}
\caption{Stage progression summary for \Name.}
\label{tab:stage_progression}
\end{table*}

Inspired by the use of structured client profiles in prior work \citep{yosef-etal-2024-assessing, wang2024clientcenteredassessmentllmtherapists}, we grounded \Name in structured profiles derived from real MI sessions between human clients and an LLM counsellor, with an original schema to support its functionalities.

\paragraph{Data Source.} 
The structured profiles are extracted from real transcripts between an automated LLM-based counsellor and real human smokers recruited online. These sessions were targeted specifically at behaviour change for smoking.  The source transcripts, as well as associated metadata, come from a previous study of a LLM-based counsellor chatbot, where real smokers recruited online remotely engaged in a conversation with the chatbot through a text-based interface, as well as completing pre- and post-conversation surveys \citep{mahmood-etal-2025-fully}. 

\paragraph{Profile Structure.}
The reference transcripts are extracted into four specific fields of the structured profiles: The \textit{Persona} field contains the general background of the client, including recent events, family relationships, occupation, religion and culture, health status, living situation, smoking history, and any other relevant background. The \textit{Motivations for Change} field includes all the potential motivations for changing their smoking behaviour that the human client expressed.  The \textit{Barriers Against Change} field records the client's expression of underlying barriers that would prevent them from changing their smoking behavior and maintain the status-quo.  Finally, the \textit{Potential Next Steps} field records next steps the client is willing to consider or adopt to support their change and reduce smoking.

\paragraph{Labeling Motivations and Barriers.}
For motivations and barriers, we apply the DARNCATs-style coding grounded in MI theory and coding practice \citep{MillerRollnick2023, CLEAR, MISC}. These categorize client language into change talk, defined as language about change, and counter-change talk (or sustain talk), defined as language maintaining the status quo. Both categories are further subdivided \citep{MillerRollnick2023} into preparatory speech (DARN: Desire, Ability, Reason, Need), mobilizing (CATs: Commitment, Activation, Taking Steps), and Other. We use the same coding scheme to classify both the motivation and barrier items into appropriate categories. As a convention, we use the character plus (+) to denote motivations or change talk items (e.g. Reason+), and the character minus (-) to denote barriers or sustain talk items (e.g. Commitment-).

\paragraph{Profile Extraction.}
We used an LLM-based profile-extraction pipeline to extract structured profiles from the transcripts. After filtering and manual review, a total of 83 distinct structured profiles were produced. Full implementation details of the pipeline, as well as an example profile, are included in Appendix \ref{app:profile_extraction}.

\subsection{Stage Progression}
\label{subsec:stage_progression}
The "Evoking" MI task begins with the counsellor meeting clients ``where they are'' \citep{MillerRollnick2023} by first eliciting their ambivalence about change, to allow them to express the current state. After that, the counsellor's goal becomes strengthening change talk, and only later moving towards actionable next steps and commitment. In consultation with MI experts, we defined three connected stages within this evoking MI task, summarized in Table~\ref{tab:stage_progression}, which the simulated client will enforce. Progression is strictly sequential: the client simulation starts in Stage~1 and can only move forward one stage at a time (Stage~1 $\rightarrow$ Stage~2 $\rightarrow$ Stage~3), with no stage skipping or backward transitions. However, the client may prevent progression at different points in the conversation under multiple conditions when sufficient progress is not achieved. This will lead to an early termination of the session from the client side.

\afterpage{%
\begin{figure*}[!t]
\centering
\includegraphics[width=0.79\textwidth]{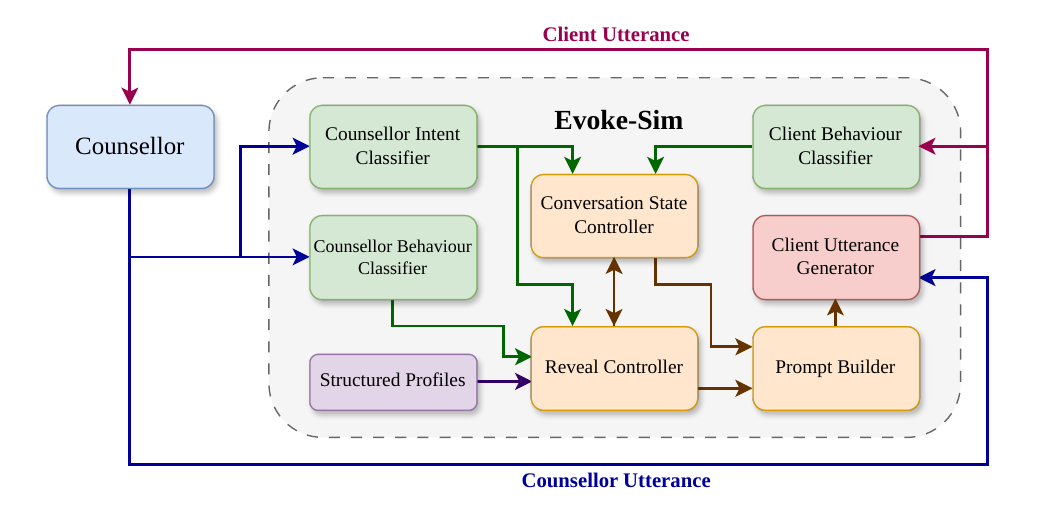}
\caption{High-level architecture and signal flow of \Name.}
\label{fig:evokesim_signal_flow}
\end{figure*}
}

\paragraph{Stage~1 (Evoking Ambivalence).}
This stage focuses on the counsellor goal of eliciting both sides of ambivalence, defined as evoking at least one motivation item and one barrier item.

\paragraph{Stage~2 (Evoking Change Talk).}
The goal for this stage is to elicit more change talk and less sustain talk. To measure the amount of client change and sustain talk, this stage maintains a sliding window \emph{change score} over a 10-turn window of recent client labels (e.g. Reason+ or Commitment-), where the score increases with each change talk utterance (labelled +) and decreases with each sustain talk utterance (-) by the client. Stage~2 transitions to Stage~3 when the change score reaches at least 3. Please refer to Appendix~\ref{app:change_score} for more details about the change score.

\paragraph{Stage~3 (Evoking Commitment to Next Steps).}
This stage focuses on evoking potential next steps for the client and eliciting their commitment language. The end goal of the conversation is satisfied when an utterance with the label Commitment+ is evoked from the client after potential next steps have been discussed.

\paragraph{Early Termination.}
In addition, if the counsellor deviates from the expected stage progression, we apply rule-based ``early-termination'' conditions across all three stages, which causes the simulated client to become unwilling to continue the conversation. In Stage~1, this occurs if the transition requirements of the stage are not met after 10 client turns. This is grounded in MI theory and practice which suggests that clients are generally ready to discuss potential next steps only after ambivalence has been elicited and sufficient preparatory change talk has emerged. This is why discussion of potential next steps (towards behavior change) is restricted to be produced by the client in Stage~3. It is also why the ``termination'' condition occurs if the counsellor seeks next steps or commitment more than three times during Stages~1 and~2. Finally, in both Stage~2 and Stage~3, we also apply a floor for the change score, derived from metadata Readiness Ruler scores introduced in Appendix~\ref{app:profile_extraction} \citep{ReadinessRuler2021}, and ``termination'' occurs when the change score reaches its floor, indicating that sustain talk has become dominant over change talk.

\subsection{Reveal Policy}
\label{subsec:reveal_policy}
A key feature of \Name that more closely aligns with real human clients is to disclose client-side information only when the counsellor actively elicits it through proper MI strategy selection and timing. We note that this feature arose because we observed empirically, in profile-grounded client simulations, it is hard to control what information gets revealed at each turn if the full profile is available from the beginning. To address this, we use a reveal policy that decides, turn by turn, what profile information is allowed to enter the conversation. This controls access to information from both sides: it limits what the counsellor can elicit from the client at each stage, and it limits what the client model can talk about because only revealed items are included in the prompt.

\paragraph{Initialization.}
At the start of each conversation, only the \emph{persona} field items from the structured profile are made visible with the client LLM prompt, while motivations, barriers, and potential next steps are hidden. Throughout the conversation, revealed items remain cumulatively visible.

\paragraph{Disclosure by Stage.}
At each turn, a reveal controller checks the current stage, what the counsellor is asking for, how the counsellor is behaving, and what has already been revealed, then makes one decision: reveal at most one new item, make no new reveal for that turn, or instruct a refusal.

Across all stages, items are only revealed if the counsellor specifically asks about them. The counsellor cannot get around this by requesting many types of items at once, as when 3 or more types of items are requested at the same turn, no new item is revealed. As additional constraints, in Stages~1 and~2, only motivations and barriers in preparatory DARNCATs categories (Desire, Ability, Reason, and Need) can be revealed, while requests for mobilizing DARNCATs categories (Activation, Commitment, and Taking Steps) as well as potential next steps are refused. In Stage~3, all profile items are allowed to be revealed. Finally, when the counsellor behaviour is not adherent to MI principles, we override the above and force either a refusal, or a sustain response grounded in visible barriers if there are any visible ones on the client side. Please refer to Appendix~\ref{app:reveal_control} for full decision details.

\subsection{\Name Architecture}
\label{subsec:implementation_flow}
Figure~\ref{fig:evokesim_signal_flow} presents a high-level visualization of how the \Name framework implements the mechanisms defined in Subsections~\ref{subsec:structured_profiles},~\ref{subsec:stage_progression}, and~\ref{subsec:reveal_policy}. The main internal modules are Counsellor Intent Classifier, Counsellor Behaviour Classifier, Client Behaviour Classifier, Conversation State Controller, Reveal Controller, Prompt Builder, and Client Utterance Generator. Among these modules, all three classifiers and the Client Utterance Generator are LLM-based, while the two controller modules and the Prompt Builder are rule-based.

Signal flow is organized as follows. A counsellor utterance enters \Name and is analyzed by the two counsellor side classifiers. Their outputs, together with client behaviour labels from the previous client turn, drive the stage update. The Reveal Controller then makes a decision on what to reveal next based on the current stage of the conversation and counsellor intent and behaviour labels. The Prompt Builder then assembles the full prompt for the next client turn  based on the current stage and the reveal decision. After client turn generation, the resulting client utterance is fed back into the Client Behaviour Classifier for the next control cycle. Full implementation details for each module are provided in Appendix~\ref{app:architecture_modules}.

%% file: experiments_results_discussion.tex
\section{Experimental Setup}
\label{sec:experimental_setup}

\newcommand{\EvokeCounsellor}{MI-Evoke\xspace}

\subsection{Research Questions}
There are two main research questions we aim to address with our experiments. Our primary research question is:
\textbf{RQ1:} Does \Name provide stronger task-aware evaluation of counsellors for the "Evoking" MI task compared to existing profile-grounded simulated clients?
We also evaluate a secondary question:
\textbf{RQ2:} Does \Name reduce generation of non-grounded statements and an overly cooperative information disclosure by the simulated client?

\subsection{Settings}
We run experiments with two client frameworks and three MI counsellors, as described below. All counsellors and clients use the same LLM, OpenAI \texttt{gpt-5.2-2025-12-11} \citep{OpenAIGPT52Docs}, with default inference parameters. We run the full factorial matrix of 2 client frameworks and 3 counsellors, resulting in 6 experimental arms. All arms are evaluated on the same set of 83 structured profiles introduced in Section~\ref{subsec:structured_profiles}.

\paragraph{Client frameworks.}
We compare two LLM-based client frameworks, both of which are profile-grounded and constrained to output one sentence per turn.
\textbf{\Name} is our proposed task-aware multi-stage client framework specifically designed to evaluate the "Evoking" MI task, which includes state progression and information revealing constraints as described in Section~\ref{subsec:implementation_flow}. The \textbf{Profile-Only} client framework serves as a baseline, and represents typical profile-grounded client simulation frameworks established in prior work \citep{yosef-etal-2024-assessing, wang2024clientcenteredassessmentllmtherapists}. The main differences are that Profile-Only clients can access the full structured profile from the start and uses only the same system prompt throughout the entire conversation, and does not receive stage-dependent instructions. Detailed prompt templates for Profile-Only clients are provided in Appendix~\ref{app:profile_only_client}.

\paragraph{MI Counsellors.}
We evaluate three LLM-based MI counsellors, all designed for the smoking cessation context:
\textbf{MINA} (MI Non-Adherent counsellor) is a baseline designed to deliberately not adhere to MI principles, prompted to be persuasive and confrontational about change (Appendix~\ref{app:mi_nonadherent_counsellor}).
\textbf{MIBot v6.3A} is a MI-adherent counsellor from a prior study \citep{mahmood-etal-2025-fully}, using a system prompt that emphasizes MI skills such as reflective listening, and eliciting ambivalence and change talk before moving to planning (Appendix~\ref{app:mibot_v63a_counsellor}).
\textbf{\EvokeCounsellor} is a counsellor designed specifically for the evoking task: it uses the same base system prompt as MIBot v6.3A, but additionally takes advantage of the same runtime state signals used by \Name and applies strategy-aligned prompting \citep{sun-etal-2025-rethinking} from a constrained MI strategy decision space (Appendix~\ref{app:mibot_evoke_counsellor}).

\subsection{Evaluation Metrics}
Because \Name explicitly models the evoking task through stage progression and controlled revealing, we can report task-specific metrics that directly evaluate how well counsellors address evoking. These metrics include ambivalence-evoking rate (fraction of conversations where ambivalence is evoked), CS$\geq$3 rate (fraction of conversations with the final change score $\geq 3$), end goal completion rate (fraction of conversations where the end goal is reached), average final stage reached (here we denote end goal reached with Stage~4 to differentiate from just entering Stage~3), average final change score, and average total conversation turns before termination. For \Name, we also report ratios of revealed items in the categories of motivations, barriers, and next steps, plus the total reveal ratio of these non-persona categories.

To complement these task-specific metrics, we also report evaluation metrics according to the MITI coding manual \citep{MITI}, which are established session-level quality measures widely adopted in prior work. MITI global scores are computed using an LLM-based approach from prior work \citep{MI-TAGS}, applying their prompt schema with the OpenAI \texttt{gpt-5.2-2025-12-11} model \citep{OpenAIGPT52Docs}, which yields 1--5 Likert scores in four dimensions: Empathy, Softening Sustain Talk, Cultivating Change Talk, and Partnership. We also report MITI behavioral coding summary measures from the Counsellor Behaviour Classifier outputs (Section~\ref{par:counsellor_behaviour_classifier}): \%CR, which is the proportion of reflections that are coded as complex reflections; R:Q, which is the reflection-to-question ratio; and total MI Non-Adherent behavior, which is the total count of "Persuade" and "Confront" behaviors.

For RQ2, we propose two metrics. First, we measure client-side non-grounded statements with a turn-level LLM judge. Here, we define non-grounded statements as generated utterances that invent facts not present in the visible set of profile items. We aggregate turn-level labels into non-grounded rate, defined as the fraction of client turns that introduce invented facts. More implementation details are provided in Appendix~\ref{app:grounding_prompts}.

Secondly, to quantify overly cooperative information disclosure, we visualize the rate of reveal for non-persona profile items by plotting the cumulative non-persona reveal fraction over turns. The plot demonstrates how quickly the client gives away hidden profile items during the conversation: a steeper early-turn slope indicates more overly cooperative behavior, while a flatter slope indicates better disclosure control. We report a 5-turn smoothed mean trajectory with 95\% confidence intervals.

\begin{table*}[!t]
\centering
\scriptsize
\setlength{\tabcolsep}{2.5pt}
\renewcommand{\arraystretch}{1.15}
\resizebox{0.99\textwidth}{!}{%
\begin{tabular}{ll|cccccc|cccc}
\thickhline
\textbf{Client} & \textbf{Counsellor} & \textbf{Ambiv.} & \textbf{CS$\geq$3} & \textbf{End Goal} & \textbf{Final Stage} & \textbf{Turns} & \textbf{Final CS} & \textbf{M-Rev} & \textbf{B-Rev} & \textbf{N-Rev} & \textbf{Total Rev} \\
\hline
Profile-Only & MINA & 0.92 & 0.90 & 0.89 & 3.71 & 26.5 & 4.7 & -- & -- & -- & -- \\
Profile-Only & MIBot v6.3A & 0.95 & 0.94 & 0.94 & 3.83 & 28.5 & 4.9 & -- & -- & -- & -- \\
Profile-Only & \EvokeCounsellor & \textbf{1.00} & \textbf{1.00} & \textbf{1.00} & \textbf{4.00} & \textbf{18.1} & \textbf{5.1} & -- & -- & -- & -- \\
\hline
\Name & MINA & 0.49 & 0.00 & 0.00 & 1.49 & 18.6 & -2.0 & 0.01 & 0.38 & 0.00 & 0.11 \\
\Name & MIBot v6.3A & \textbf{1.00} & 0.27 & 0.23 & 2.49 & 33.5 & -0.9 & 0.32 & \textbf{0.90} & 0.13 & 0.45 \\
\Name & \EvokeCounsellor & \textbf{1.00} & \textbf{0.99} & \textbf{0.99} & \textbf{3.98} & \textbf{18.5} & \textbf{5.1} & \textbf{0.47} & 0.49 & \textbf{0.56} & \textbf{0.46} \\
\thickhline
\end{tabular}
}
\caption{Task-aware evaluation metrics by client framework and counsellor. Abbreviations: Ambiv. = ambivalence-evoking rate (closer to 1 is better), CS$\geq$3 = rate of conversations with final change score $\geq 3$ (closer to 1 is better), End Goal = end goal completion rate (closer to 1 is better), Average Final Stage (closer to 4 is better) 1; Average Number of Turns, Average Final CS = final change score, M-Rev/B-Rev/N-Rev = reveal ratio for motivations/barriers/next steps, Total Rev = total reveal ratio for all non-persona categories.}
\label{tab:task_aware_results}
\end{table*}

\begin{table*}[!t]
\centering
\scriptsize
\setlength{\tabcolsep}{2.5pt}
\renewcommand{\arraystretch}{1.15}
\resizebox{0.75\textwidth}{!}{%
\begin{tabular}{ll|ccccc|ccc}
\thickhline
\textbf{Client} & \textbf{Counsellor} & \textbf{Global} & \textbf{CCT} & \textbf{SST} & \textbf{Part.} & \textbf{Emp.} & \textbf{\%CR} & \textbf{R:Q} & \textbf{MI NonAdh.} \\
\hline
Profile-Only & MINA & 2.19 & 3.94 & 1.88 & 1.01 & 1.93 & 64.6 & 0.26 & 18.5 \\
Profile-Only & MIBot v6.3A & \textbf{4.70} & \textbf{4.93} & \textbf{4.47} & \textbf{4.71} & \textbf{4.69} & \textbf{94.8} & \textbf{0.82} & 0.20 \\
Profile-Only & \EvokeCounsellor & 4.44 & 4.90 & 4.33 & 4.22 & 4.31 & 89.5 & 0.56 & \textbf{0.02} \\
\hline
\Name & MINA & 1.05 & 1.06 & 1.00 & 1.00 & 1.16 & 68.3 & 0.17 & 12.5 \\
\Name & MIBot v6.3A & 3.81 & 3.90 & 2.86 & 4.04 & \textbf{4.45} & \textbf{96.4} & \textbf{0.90} & 0.05 \\
\Name & \EvokeCounsellor & \textbf{4.37} & \textbf{4.94} & \textbf{4.19} & \textbf{4.19} & 4.16 & 85.9 & 0.63 & \textbf{0.01} \\
\thickhline
\end{tabular}
}
\caption{MITI metrics by client framework and counsellor. Abbreviations: Global = overall MITI global score, CCT = Cultivating Change Talk, SST = Softening Sustain Talk, Part. = Partnership, Emp. = Empathy, \%CR = percent complex reflections, R:Q = reflection-to-question ratio, MI NonAdh. = total MI Non-Adherent behaviors.}
\label{tab:miti_results}
\end{table*}

\afterpage{\afterpage{%
\begin{figure*}[!t]
\centering
\includegraphics[width=0.35\textwidth]{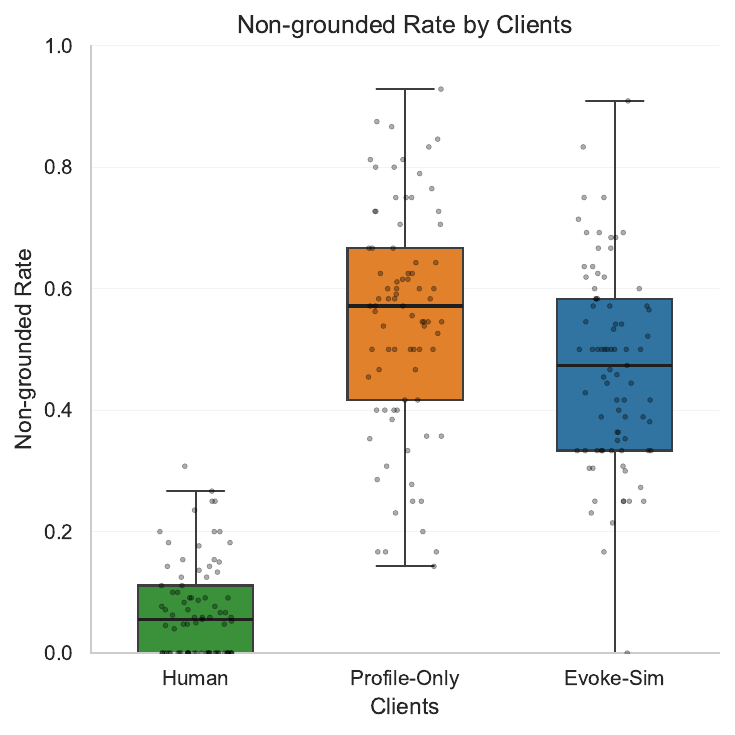}
\hfill
\includegraphics[width=0.61\textwidth]{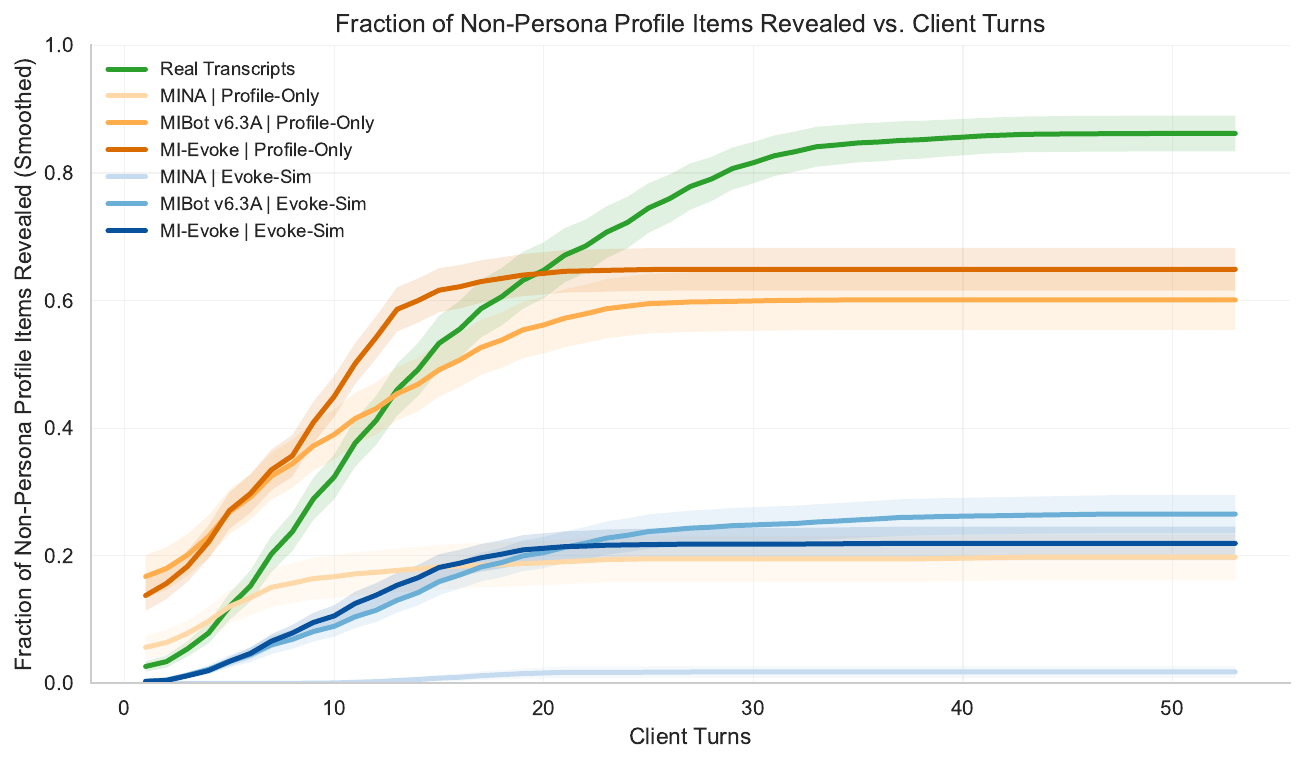}
\caption{Left: distributions of non-grounded rate in conversations between MIBot v6.3A and human clients, Profile-Only clients, and \Name clients. Right: smoothed mean fraction of non-persona profile items revealed across client turns, with 95\% Confidence Interval (CI) for the mean.}
\label{fig:secondary_rq}
\end{figure*}
}}

\section{Results and Discussion}
\label{sec:results}

\subsection{RQ1}
Tables~\ref{tab:task_aware_results} and~\ref{tab:miti_results} report the primary results across all six experimental arms (statistical tests in Appendix~\ref{app:omnibus_tests}). The first table focuses on metrics specific to the "Evoking" MI task, while the second table reports MITI global and summary measures. 

Under Profile-Only, there is not a clear separation in performance between the counsellors across all task-aware metrics. All three counsellors achieve high task-outcome scores, including MINA. Although \EvokeCounsellor is the best performer, end goal completion rate remains high across all counsellors (0.89 for MINA, 0.94 for MIBot v6.3A, and 1.00 for \EvokeCounsellor). However, under \Name, the same counsellors are clearly separated by the same metrics. MINA fails to reach the end goal under \Name, with less than half of conversations successfully evoking ambivalence. \EvokeCounsellor reaches near-ceiling end goal completion with strongly positive Final CS. MIBot v6.3A remains substantially better than MINA, but remains far below \EvokeCounsellor on end goal completion and stage progression. These results suggest that \textbf{\Name is much better at differentiating MI counselling quality in terms of how well they perform in the evoking task.}

An additional benefit of \Name is that each \textbf{task-aware metric offers uniquely valuable insights for the evaluation of the MI counsellors at the evoking task.} Stage-specific metrics show that Stage~1 ambivalence elicitation separates MINA from two MI-adherent counsellors, while Stage~2 change score threshold separates MIBot v6.3A from \EvokeCounsellor. In addition, the large gap in final change score under \Name suggests that a generic MI system prompt alone is not sufficiently strategic for the evoking task. Average turns captures efficiency when viewed together with end goal rate: across both frameworks, \EvokeCounsellor achieves a very high end goal rate with the fewest turns on average (Profile-Only: End Goal = 1.00, Turns = 18.1; \Name: End Goal = 0.99, Turns = 18.5), highlighting its competence and efficiency, whereas MINA’s short conversations (Turns = 18.6) under \Name are largely explained by early termination (End Goal = 0.00; Final Stage = 1.49) rather than effective evoking. The reveal metrics add another layer of interpretation: in our results, stronger counsellors such as \EvokeCounsellor are associated with higher motivation and next-step reveals (M-Rev 0.47, N-Rev 0.56), compared with a weak counsellor like MINA (M-Rev 0.01, N-Rev 0.00). The high barrier reveal ratio of MIBot v6.3A (B-Rev 0.90) suggests that it evokes barriers too often, limiting its ability to navigate the conversation towards client change talk. 

Compared with the task-aware results above, MITI metrics mainly show a large gap between MINA and the other counsellors, while providing limited separation between MIBot v6.3A and \EvokeCounsellor. In Profile-Only, MINA is clearly the lowest across global and summary measures, while MIBot v6.3A and \EvokeCounsellor are close across most metrics, with MIBot v6.3A generally stronger. This suggests that \textbf{under typical profile-based client simulation frameworks, automated MITI-based metrics do not tell the full story, especially not for the evoking task.} Interestingly, under \Name, although MINA drops further and remains clearly lowest, the top global MITI performance shifts to \EvokeCounsellor (Global, CCT, SST, and Partnership), while MIBot v6.3A stays close and higher on Empathy and on the summary measures (\%CR and R:Q). This suggests that \textbf{\Name makes MITI global scoring more sensitive to task performance}, while also showing why MITI metrics alone are insufficient. It is also important to note that, unlike the global scores, summary measures such as \%CR and R:Q may not be strictly monotonic indicators of better counselling in practice, and should be examined in a case-by-case manner. 

\subsection{RQ2}

Figure~\ref{fig:secondary_rq} (left) compares distributions of non-grounded rate across both client systems with the counsellor fixed to MIBot v6.3A. In addition, they are compared to the real transcripts as a baseline, which come from sessions between online-recruited human clients and the same MIBot v6.3A counsellor. Mean non-grounded rates are 0.070 for human clients, 0.55 for Profile-Only, and 0.47 for \Name. Relative to Profile-Only, \textbf{\Name reduces client-side non-grounded statements under matched counsellor conditions}, while still remaining above the real-transcript baseline. This suggests that \Name's control on visibility of profile items improves clients' ability to stay grounded to the structured profiles, but does not fully close the gap yet. Note that comparison implicitly treats real clients as grounded to a latent true profile, which may not always hold in practice as real clients can also introduce spontaneous or inconsistent facts in counselling sessions. The complete boxplot is included in Appendix~\ref{app:grounding_all_settings}.

Figure~\ref{fig:secondary_rq} (right) shows the rate of reveal over client turns using a 5-turn smoothed trajectory. The MINA counsellor consistently causes a lot fewer reveals compared to other counsellors in both client frameworks. Looking at the conversations with the two MI-adherent counsellors, \Name clients generally reveal non-persona information more slowly and to a lower final level than Profile-Only clients. Notably, Profile-Only conversations tend to share a large fraction of non-persona profile items within the first few turns (as many as 20\%), which is much higher than real transcripts, while \Name conversations show more gradual and controlled information disclosure over turns. This shows that \textbf{\Name is overall less cooperative in information disclosure, especially in the early turns.} Statistical tests for all RQ2 related results are included in Appendix~\ref{app:rq2_pairwise_tests}.

%% file: conclusions.tex
\section{Conclusion}
\label{sec:conclusions}

We propose \Name, a task-aware multi-stage client simulation framework for evaluating MI counsellors on the evoking MI task. Results show that \Name provides stronger task-aware differentiation of counsellor quality than existing profile-grounded clients, while also making MITI global scores more aligned with task-specific performance. Additionally, \Name is shown to reduce client-side non-grounded statements and slow information disclosure. Overall, \Name sets a higher standard for LLM-based MI counsellors by holding them to task-specific performance in evoking.

%% file: limitations.tex
\section{Limitations}
\label{sec:limitations}

This study has several limitations. Although \Name is a task-aware client simulation framework, its scoped is limited to only the "Evoking" MI task. As a result, its stage logic, reveal policy, and task-aware metrics are only designed for evoking and may not transfer directly to other MI tasks (e.g., Engaging, Focusing, and Planning) without additional design and verifications.

The source transcripts used for profile extraction, and often used as comparison references, are full MI sessions that can extend beyond evoking. This creates a potential scope mismatch between our evoking-specific simulation framework and the reference transcripts. In addition, our reveal-based metrics may have a limitation: the non-visible profile items were extracted from those same transcripts, and we treat that extracted set as the full set of revealable information. That assumption may be incomplete, because the real underlying client profiles can contain additional information not captured by extraction.

Although the evaluation criteria enabled by \Name are based on MI principles in general, all experiments are only in the smoking-cessation context, and in a text-only interface. Generalization to other target behavior changes, such as reducing alcohol consumption and weight management, as well as MI sessions in other modalities such as video or audio, all remain as open questions yet to be explored.

Our partition of the evoking task into three stages was developed through collaboration with MI experts, but alternative expert interpretations are possible. In addition, nuanced strategy choices used by specific expert counsellors may not yet be fully captured by the current state and reveal design in \Name.

Several internal modules of \Name rely on an LLM-based methodology, including profile extraction, automated behavior and intent classification, and MITI global scoring. Although many such LLM-based methodology have already been established and properly verified by prior work, these modules can still introduce inaccuracies. They do not replace expert human coding and should not be interpreted as direct evidence of clinical effectiveness.

Finally, this work evaluates counsellor quality in simulation rather than direct patient outcomes. Strong performance under \Name indicates robustness under controlled simulation settings, but additional studies are needed to establish predictive validity for counselling systems aimed at human clients.

%% file: ethical_considerations.tex
\section{Ethical Considerations}
\label{sec:ethical_considerations}

\Name proposes a new angle for building client simulation frameworks that focuses on task-aware evaluation of MI quality. Rather than evaluating counsellors with only general MI quality scores and behavior analysis, we propose a shift towards focusing on how well the counsellors are performing with respect to specific MI tasks, starting with evoking. We encourage future work to report task-aware evaluation metrics alongside standard MI metrics, extend the client simulation framework to support additional MI tasks, and validate these benchmarks with expert human review and potentially studies with human participants.

This work also carries potential risks. \Name could be misused to generate realistic synthetic counselling clients outside of research settings, including for deceptive demonstrations or for overclaiming the readiness of automated MI counsellors without thoughtful human verifications. Furthermore, \Name only focuses on the evoking task in the context of smoking cessation, which could encapsulate narrow assumptions about what counts as high quality MI counselling. We therefore position \Name as a research benchmark rather than a deployment tool, and we encourage future use to include explicit reporting of domain and task scope, as well as careful expert review before any human-facing application.

The \Name framework is grounded in transcripts with human participants through profile extraction. All data used in this paper come from a publicly available dataset released by a prior study in which participants consented to use of their data for research purposes. The source transcripts are distributed under a Creative Commons Attribution-ShareAlike 4.0 license (CC BY-SA 4.0), and our use of them to derive structured profiles is intended to remain consistent with those reuse conditions. The source data were anonymized before use. In preparing our released resources, we manually checked for personally identifying information as well as offensive content and did not identify any.

%% file: profile_extraction_multicolumn.tex
\section{Profile Extraction Pipeline}
\label{app:profile_extraction_multicolumn}
\label{app:profile_extraction}

\begin{multicols}{2}
We run an LLM-based profile-extraction pipeline over the smoking-cessation counselling transcripts described in Section~\ref{sec:methodology}. Starting from 106 eligible transcripts, we use an LLM to extract one candidate structured profile per conversation, apply LLM-based filtering and manual review, and then apply deterministic post-processing to incorporate selected metadata. After this process, we were left with 83 distinct structured profiles.

\paragraph{Initial Extraction.}
Client profiles were first extracted using OpenAI's \texttt{gpt-5.2-2025-12-11} \citep{OpenAIGPT52Docs} in a zero-shot fashion using default parameters, given one reference transcript at a time and requiring JSON formated outputs. The full extraction prompt is shown in \hyperref[box:profile_extraction_prompt_multicolumn]{Box~A.1}. Instructions were provided to classify motivation and barrier items into DARNCATs labels as part of the extraction process. The exact definitions for which are provided in Table~\ref{tab:darncats_labels_multicolumn}. We then filtered the extracted candidates to keep only profiles that had all four required sections (\textit{persona}, \textit{motivations for change}, \textit{barriers against change}, and \textit{potential next steps}) non-empty, and contained at least one non-CATs motivation item and one non-CATs barrier item, where CATs refers to Commitment, Activation, and Taking Steps.

\paragraph{Overlap Pruning.}
The information revealing mechanism detailed in Section~\ref{sec:methodology} requires that only the persona items are initially visible to the simulated client, while the motivations, barriers, and potential next steps are hidden until the revealed later. To ensure that the persona section of the profile does not contain information that overlaps with the other sections, we apply an LLM-based overlap detection step followed by rule-based pruning, using the same OpenAI model \citep{OpenAIGPT52Docs} in a similar fashion. The full overlap-detection prompt is shown in \hyperref[box:overlap_detection_prompt_multicolumn]{Box~A.2}. In the rule-based pruning step that follows, any persona item having any overlaps with any other section is removed.

After both stages of LLM-based processing, samples of the outputs are manually reviewed by three of the co-authors to check for quality and consistency, and to identify any systematic issues. In both cases, the LLM-based outputs were found to be closely aligning with human standards.

\paragraph{Incorperating Metadata in the Profiles.}
For each conversation, we select six metadata attributes and translate them into natural-language persona statements using fixed templates: daily cigarette count, time to first cigarette after waking, any quit attempt in the prior week, and three items from the Readiness Rulers: confidence in quitting, importance of quitting, and readiness to quit. The Readiness Rulers are brief self-report MI measures where the clients place themselves on a 0--10 continuum \citep{ReadinessRuler2021}. These metadata-derived statements make the first six persona items for every structured profile. The conversion logic is rule-based and deterministic, so the same metadata input always yields the same profile text. Table~\ref{tab:meta_to_persona_rules_multicolumn} summarizes the field-level mapping, and Table~\ref{tab:ruler_bucket_map_multicolumn} summarizes the bucket mapping used for the confidence, importance, and readiness ruler inputs. An example of a final structured client profile after all processing steps is shown in \hyperref[box:profile_example_multicolumn]{Box~A.3}.
\begin{table*}[t]
\centering
\small
\begingroup
\renewcommand{\arraystretch}{1.30}
\begin{tabular}{p{0.18\linewidth}p{0.76\linewidth}}
\thickhline
\textbf{Label} & \textbf{Definition} \\
\hline
\multicolumn{2}{l}{\textbf{Motivations for Change -- Preparatory (DARN)}} \\
Desire+ & A special type of reason stating the client's willingness to alter the target behavior. \\
Ability+ & A statement indicating that the client is able to change. \\
Reason+ & A statement indicating a rationale for changing the target behavior. \\
Need+ & A special type of reason stating the client's need to change. \\[1.5ex]
\multicolumn{2}{l}{\textbf{Motivations for Change -- Mobilizing (CATs)}} \\
Commitment+ & A statement that the client will change, or an idea for how the client could change. \\
Activation+ & A statement that the client is leaning towards action and open to change, but has not yet made a clear commitment. \\
TakingSteps+ & A statement that the client has already begun to change; this represents steps taken in the recent past (within approximately the past week). \\[1.5ex]
\multicolumn{2}{l}{\textbf{Motivations for Change -- Other}} \\
Other+ & Any other statement about changing the target behavior. Includes hypothetical situations or circumstances that would convince the client to change, and problem recognition. \\[1.5ex]
\multicolumn{2}{l}{\textbf{Barriers against Change -- Preparatory (DARN)}} \\
Desire- & A special type of reason, expressing the client's unwillingness to change or wish to partake in the target behavior. \\
Ability- & A statement that client is unable or unconfident about change. \\
Reason- & A statement indicating a rationale for not changing or for why change is unnecessary. \\
Need- & A special type of reason stating a need not to change or to stay the same. \\[1.5ex]
\multicolumn{2}{l}{\textbf{Barriers against Change -- Mobilizing (CATs)}} \\
Commitment- & A statement that the client will not change, or an idea for how not to change/to stay the same. \\
Activation- & A statement that the client is leaning towards not making changes or not open to change, but has not committed to not changing just yet. \\
TakingSteps- & A statement that the client is already resisting change; this represents steps taken in the recent past (within approximately the past week). \\[1.5ex]
\multicolumn{2}{l}{\textbf{Barriers against Change -- Other}} \\
Other- & A statement that is clearly counter-change talk but does not fit reasonably into the other categories. This includes minimization of problems and hypothetical statements about non-change. \\
\thickhline
\end{tabular}
\endgroup
\caption{DARNCATs labels used to classify motivation and barrier profile items.}
\label{tab:darncats_labels_multicolumn}
\label{tab:darncats_labels}
\end{table*}

\begingroup
\setlength{\floatsep}{4pt}
\setlength{\textfloatsep}{4pt}
\begin{table*}[!t]
\centering
\small
\renewcommand{\arraystretch}{1.40}
\begin{tabular}{p{0.22\linewidth}p{0.28\linewidth}p{0.34\linewidth}}
\thickhline
\rule{0pt}{2.5ex}\textbf{Metadata Field} & \textbf{Allowed Values} & \textbf{Converted Persona Items} \\[0.3ex]
\hline
Daily cigarette count & Integer $n \ge 0$ & ``The client smokes $n$ cigarettes per day.'' (singular when $n=1$). For $n=0$: ``The client does not smoke cigarettes (FLAG: Unexpected for this study).'' \\
Time to first cigarette after waking & One of: ``within 5 minutes'', ``6 to 30 minutes'', ``31 to 60 minutes'', ``after 60 minutes'' & ``The client smokes their first cigarette \ldots after waking up.'' \\
Any quit attempt in the prior week & Boolean value: True/False & True: ``The client made at least one attempt to quit smoking in the week before the conversation.'' False: ``The client did not make an attempt to quit smoking in the week before the conversation.'' \\
Confidence in quitting (0--10) & An integer score in $[0,10]$ & ``The client thinks that they are [bucket] confident that they would succeed at stopping smoking if they start now.'' \\
Importance of quitting (0--10) & An integer score in $[0,10]$ & ``The client thinks that it is [bucket] important for them to stop smoking right now.'' \\
Readiness to quit (0--10) & An integer score in $[0,10]$ & ``The client thinks that they are [bucket] ready to start making a change at stopping smoking right now.'' \\
\thickhline
\end{tabular}
\caption{Metadata fields and deterministic conversion rules for persona items.}
\label{tab:meta_to_persona_rules_multicolumn}
\label{tab:meta_to_persona_rules}

\vspace{20pt}
\begin{tabular}{ll}
\thickhline
Ruler Score & Bucket Phrase \\
\hline
0 & not at all \\
1--3 & not very \\
4--6 & moderately \\
7--9 & very \\
10 & extremely \\
\thickhline
\end{tabular}
\caption{Bucket mapping for confidence, importance, and readiness metadata fields.}
\label{tab:ruler_bucket_map_multicolumn}
\label{tab:ruler_bucket_map}
\end{table*}
\endgroup

\end{multicols}
\onecolumn
\nolinenumbers

\newtcolorbox{promptbox}[1]{
    breakable,
    colback=blue!4,
    colframe=blue!65!black,
    width=\textwidth,
    fontupper=\small,
    title={#1}
}

\phantomsection
\label{box:profile_extraction_prompt_multicolumn}
\label{box:profile_extraction_prompt}
\begin{promptbox}{Box A.1: Full Prompt Used to Extract Client Profiles}
You are an assistant that analyzes real counselling transcripts and extracts a concise, structured profile of the client. The context of these counselling transcripts is smoking cessation. Base everything ONLY on what the client actually says in the transcript. Do not invent or assume facts that are not clearly supported by the client's utterances. Write in clear, concise third-person statements that begin with 'The client...'.

\#\# Task

Your task is to identify the client's profile based on the provided counseling conversation. Focus on the following aspects:

\begin{itemize}
\item **Persona**: Describe the client in a bullet-style list of short, third-person statements (for example, "The client lives with..."), covering personal details such as recent events, family relationships, occupation, religion and culture, health status, living situation, smoking history, and any other relevant background. Each item in the persona list should be a concise third-person description of the client, not a copied sentence from the transcript and not a direct quote.
\item **Motivations for Change**: Explain why the client wants to change this behavior (due to family, health, work, etc.), and extract any client statements that move towards changing smoking behavior. For each such statement, classify it using DARNCATs change talk categories and summarize the idea in your own concise third-person words (for example, "The client wants to quit smoking to improve their health"), rather than copying or quoting the client's exact utterances.
\item **Barriers Against Change**: Detail the client's inner beliefs or statements that move away from change or towards sustaining smoking (counter-change talk). For each such statement, classify it using DARNCATs counter-change talk categories and summarize the idea in your own concise third-person words (for example, "The client believes they need cigarettes to manage stress"), rather than copying or quoting the client's exact utterances.
\item **Potential Next Steps**: Describe any plans the client is willing to adopt or consider, like reducing frequency, altering the environment, or seeking help. Use one sentence for each plan.
\end{itemize}

\#\# DARNCATs Definitions and Examples (adapted to smoking)

Counter-change talk. This type of client language refers to any movement away from change, or towards sustaining the target behavior. Note that "change" here is defined in reference to the target behavior. Within the context of treatment for problem smoking, for example, counter-change talk is coded in relation to maintaining or increasing smoking behavior. Clients may express counter-change talk on other subjects (e.g., change in a relationship, moving to a new apartment), but these are not coded unless directly related to the identified target behavior change. Counter-change talk need not have an oppositional quality nor an emotional charge. The key is that the client language favors not changing the target behavior, representing status quo or movement backward. Endorsing or expressing agreement with counter-change talk offered by the therapist should be coded as an instance of counter-change talk. Each different counter-change talk statement counts as one instance of counter-change talk. For example, if a client lists several different reasons against or disadvantages of change, each one is coded as counter-change talk (e.g., a volley that included a Desire-, Need-, and Other- would count as three counter-change talk tallies, and a string of four Reason-'s would count as four counter-change talk tallies).

Some sub-categories of counter-change talk include:
\begin{itemize}
\item Reason-: A statement indicating a rationale for not changing or for why change is unnecessary.
\begin{itemize}
\item "Going out with friends wouldn't be any fun without having a smoke together."
\item "The kids stress me out too much when I'm not smoking."
\item "My health seems fine, so I don't see a problem with my smoking."
\end{itemize}
\item Desire-: A special type of reason, expressing the client's unwillingness to change or wish to partake in the target behavior.
\begin{itemize}
\item "If I could, I would smoke every day until I'm 90."
\item "I love smoking."
\end{itemize}
\item Need-: A special type of reason stating a need not to change or to stay the same.
\begin{itemize}
\item "Treatment for smoking isn't something that I need right now."
\item "I don't need to quit smoking entirely."
\item "I need to keep smoking if I want to keep these friends."
\end{itemize}
\item Ability-: A statement that client is unable or unconfident about change.
\begin{itemize}
\item "It's just too hard to change my smoking after so many years."
\item "I'm feeling pretty low on the confidence scale about quitting."
\end{itemize}
\item Commitment-: A statement that the client will not change, or an idea for how not to change/to stay the same.
\begin{itemize}
\item "As soon as I get out of this program I'm going to buy another pack."
\item "I'm not going to say that I won't smoke ever again."
\end{itemize}
\item Activation-: A statement that the client is leaning towards not making changes or not open to change, but has not committed to not changing just yet.
\begin{itemize}
\item "I'm not willing to try what you just suggested"
\item "I would not consider quitting at least for this month."
\item "I'm not really thinking about smoking less."
\end{itemize}
\item Taking Steps-: A statement that the client is already resisting change; this represents steps taken in the recent past (within approximately the past week).
\begin{itemize}
\item "I picked up another carton of cigarettes over the weekend."
\item "I stopped going to my smoke-free support group."
\end{itemize}
\item Other-: A statement that is clearly counter-change talk but does not fit reasonably into the other categories. This includes minimization of problems and hypothetical statements about non-change.
\begin{itemize}
\item "Another warning from my doctor about smoking isn't that big of a deal to me."
\item "If I were allowed to smoke at work, I'd light up right at my desk."
\end{itemize}
\end{itemize}

Change talk. This type of client language refers to any movement towards change or away from the target behavior. As with counter-change talk, "change" here is defined specifically in reference to the target behavior. The client makes a statement that directly or indirectly shows evidence of at least one of the following categories, which have the quality of moving forward in the direction of change in the target behavior. Within the context of treatment for problem smoking, for example, change talk is coded in reference to reducing or stopping smoking behavior. Each different change talk statement counts as one instance of change talk. For example, if a client lists several different reasons for or advantages of change, each one is coded as change talk. As with counter-change talk, endorsing or expressing agreement with change talk offered by the therapist should be coded as an instance of change talk.

Some sub-categories of change talk include:
\begin{itemize}
\item Reason+: A statement indicating a rationale for changing the target behavior.
\begin{itemize}
\item "Quitting smoking would help me feel less distracted at work."
\item "I hate the way my lungs feel after I smoke."
\item "My family needs me to be present at home, not outside smoking."
\end{itemize}
\item Desire+: A special type of reason stating the client's willingness to alter the target behavior.
\begin{itemize}
\item "I really want to get started with quitting smoking."
\item "I don't even feel like having a cigarette today."
\end{itemize}
\item Need+: A special type of reason stating the client's need to change.
\begin{itemize}
\item "I have to do this and quit smoking."
\item "Getting help to stop smoking is what I need right now."
\end{itemize}
\item Ability+: A statement indicating that the client is able to change.
\begin{itemize}
\item "I know that I can quit if I try hard enough."
\item "This doesn't seem so difficult if I take it one day at a time."
\end{itemize}
\item Commitment+: A statement that the client will change, or an idea for how the client could change.
\begin{itemize}
\item "I'll do whatever it takes to cut down on my smoking."
\item "I could start by tossing out all the cigarettes at home."
\end{itemize}
\item Activation+: A statement that the client is leaning towards action and open to change, but has not yet made a clear commitment.
\begin{itemize}
\item "I'm willing to give it another try."
\item "I would consider quitting this month."
\item "I'm thinking about reducing the amount I smoke."
\end{itemize}
\item Taking Steps+: A statement that the client has already begun to change; this represents steps taken in the recent past (within approximately the past week).
\begin{itemize}
\item "At dinner last night I told my family that I'm going to quit smoking."
\item "I've already cut down the number of cigarettes I smoke this week."
\end{itemize}
\item Other+: Any other statement about changing the target behavior. Includes hypothetical situations or circumstances that would convince the client to change, and problem recognition.
\begin{itemize}
\item "My smoking is out of control."
\item "If I could move to a smoke-free environment, I'd be less likely to feel the urge to smoke."
\end{itemize}
\end{itemize}

\#\# DARNCATs Label Set for Coding

When you classify motivations for change or barriers against change, use ONLY the following category labels:
\begin{itemize}
\item Reason+, Desire+, Need+, Ability+, Commitment+, Activation+, TakingSteps+, Other+
\item Reason-, Desire-, Need-, Ability-, Commitment-, Activation-, TakingSteps-, Other-
\end{itemize}

Base all categorizations only on what the client actually says about smoking. If a statement is clearly about changing smoking, it should appear in **Motivations for Change** with a "+" category. If a statement is clearly about keeping smoking the same or against change, it should appear in **Barriers Against Change** with a "-" category.

\#\# Output Format

Return the client profile in JSON format with the following structure:
\begin{lstlisting}[basicstyle=\ttfamily\small]
{
  "persona": [
    "The client lives with ...",
    "The client works as ...",
    ...
  ],
  "motivations_for_change": [
    { "category": "Reason+", "text": "..." },
    { "category": "Desire+", "text": "..." },
    ...
  ],
  "barriers_against_change": [
    { "category": "Reason-", "text": "..." },
    { "category": "Ability-", "text": "..." },
    ...
  ],
  "potential_next_steps": [
    "...",
    "..."
  ]
}
\end{lstlisting}

NOTE:
\begin{itemize}
\item All keys must match the names shown above exactly.
\item There should never be two items covering the exact same fact about the client, and that applies to items both in the same category and across categories. If something belongs more in **Motivations for Change**, **Barriers Against Change** or **Potential Next Steps**, it should not be included in **Persona**.
\item If there is no information for \texttt{motivations\_for\_change}, \texttt{barriers\_against\_change}, or \texttt{potential\_next\_steps} in the conversation, set that field's value to the exact string "None" instead of using an empty list, null, or omitting the field.
\item If there is no persona information at all in the conversation, set the value of \texttt{persona} to the exact string "None" instead of using an empty list, null, or omitting the field.
\item You must respond with a single valid JSON object only, with no explanation, no commentary, and no Markdown code fences. Do not write \verb|```json| or any text before or after the JSON.
\end{itemize}

\#\# Given Conversation

\texttt{\{...\}}

\#\# Instruction

Based on the conversation, provide the client's profile in JSON format following the schema above, using DARNCATs categories for all \texttt{motivations\_for\_change} and \texttt{barriers\_against\_change} entries.
\end{promptbox}

\phantomsection
\label{box:overlap_detection_prompt_multicolumn}
\label{box:overlap_detection_prompt}
\begin{promptbox}{Box A.2: Full Prompt Used to Detect Overlaps Between Persona and Other Profile Sections}
You are a precise assistant that detects overlapping or redundant information across sections of a client profile. Treat "overlap" as meaning the persona item explains the same fact or claim as an item from other categories, even if wording differs. Only mark overlaps that are clearly the same statements or sharing the same exact information. If either the persona item or the other item in question (motivations, barriers, next steps) adds any new information, potentially because of a different phrasing, do NOT mark it as overlapping. If uncertain, leave it out. Use the transcript as grounding to check meaning consistency.

\begin{itemize}
\item Profile number: \texttt{\{...\}}
\item Conversation ID: \texttt{\{...\}}
\item Transcript:
\begin{lstlisting}[basicstyle=\ttfamily\small]
----------------
{...}
----------------
\end{lstlisting}
\item Persona items (numbered as P\#): \texttt{\{...\}}
\item Motivations for change (numbered as M\#): \texttt{\{...\}}
\item Barriers against change (numbered as B\#): \texttt{\{...\}}
\item Potential next steps (numbered as N\#): \texttt{\{...\}}
\end{itemize}

Return a JSON object ONLY with this exact shape:
\begin{lstlisting}[basicstyle=\ttfamily\small]
{
  "overlaps": {
     "<persona_id>": [
        "<overlapping_item_id>"
     ],
     ... repeat for each persona item ...
  }
}
\end{lstlisting}

\begin{itemize}
\item Include every persona item as a key, even if it has zero overlaps (use empty list []).
\item Use ONLY the IDs (P\#, M\#, B\#, N\#) in the JSON. Do NOT include the text.
\item Do not invent new persona items.
\item Output JSON only, no markdown.
\end{itemize}
\end{promptbox}

\clearpage
\phantomsection
\label{box:profile_example_multicolumn}
\label{box:profile_example}
\begin{tcolorbox}[
    breakable,
    colback=green!6,
    colframe=green!55!black,
    width=\linewidth,
    fontupper=\small,
    title={Box A.3: Example Structured Client Profile}
]
\textbf{Persona}
\begin{itemize}
  \item The client smokes 10 cigarettes per day.
  \item The client smokes their first cigarette 6 to 30 minutes after waking up.
  \item The client made at least one attempt to quit smoking in the week before the conversation.
  \item The client thinks that they are not very confident that they would succeed at stopping smoking if they start now.
  \item The client thinks that it is very important for them to stop smoking right now.
  \item The client thinks that they are moderately ready to start making a change at stopping smoking right now.
  \item The client is experiencing stress related to work and caring for children.
\end{itemize}

\textbf{Motivations for Change}
\begin{itemize}
  \item \textbf{Reason+}: The client wants to improve breathing and long-term health by reducing smoking.
  \item \textbf{Activation+}: The client is willing to try reducing smoking during selected daily breaks.
  \item \textbf{Desire+}: The client wants to quit smoking and be a healthier parent.
\end{itemize}

\textbf{Barriers Against Change}
\begin{itemize}
  \item \textbf{Reason-}: The client experiences smoking as immediate relief from stress.
  \item \textbf{Ability-}: The client reports low confidence in sustaining a quit attempt.
  \item \textbf{Activation-}: The client is currently inclined to keep smoking in high-stress moments.
\end{itemize}

\textbf{Potential Next Steps}
\begin{itemize}
  \item The client is willing to start with one smoke-free break each day.
  \item The client is willing to replace one smoking break with a short walk or breathing exercise.
\end{itemize}
\end{tcolorbox}

%% file: change_score_multicolumn.tex
\section{Change Score Computation}
\label{app:change_score_multicolumn}
\label{app:change_score}

\begin{multicols}{2}
This appendix gives the exact change score computation used by the stage manager in \Name.

\paragraph{Scope.}
Change score tracking starts in Stage~2 and continues to be maintained in Stage~3. Stage~1 only tracks ambivalence and does not update the change score.

\paragraph{Rolling window.}
At client turn $t$ in Stage~2 and Stage~3, let $W_{\mathrm{C}}^{(t)}$ denote the rolling window of the most recent up to 10 client behaviour labels. Let $C_{+}$ denote all plus DARNCATs labels (change talk categories) and $C_{-}$ denote all minus DARNCATs labels (sustain talk categories), following Table~\ref{tab:darncats_labels_multicolumn} in Appendix~\ref{app:profile_extraction_multicolumn}.
\[
\begin{aligned}
\#C_{+}^{(t)}
&=
\left|
\left\{
y\in W_{\mathrm{C}}^{(t)}
\;\middle|\;
y\in C_{+}
\right\}
\right|
\end{aligned}
\]
\[
\begin{aligned}
\#C_{-}^{(t)}
&=
\left|
\left\{
y\in W_{\mathrm{C}}^{(t)}
\;\middle|\;
y\in C_{-}
\right\}
\right|
\end{aligned}
\]
The change score at turn $t$ is then:
\[
s^{(t)}=\#C_{+}^{(t)}-\#C_{-}^{(t)}
\]

\paragraph{Profile-conditioned floor.}
The floor for the change score is unique to each client profile. It is initialized from the three Readiness Ruler scores from client metadata, introduced in Appendix~\ref{app:profile_extraction_multicolumn} \citep{ReadinessRuler2021}. Let $c$, $i$, and $r$ denote the confidence, importance, and readiness ruler scores, respectively:
\[
c,i,r \in \{0,1,\ldots,10\}
\]
Note that, in the source study from which these profiles are extracted, participants were required to be either low-confidence ($c \le 5$), or high-confidence but discordant ($c > 5$ and $c-i>5$) \citep{andrewMI}.
The sum of $c$, $i$, and $r$ gives the total ruler score $R$:
\[
R=c+i+r
\]
From this total, we compute a clipped bucket index $k$:
\[
k=\operatorname{clip}\!\left(\left\lfloor\frac{R-4}{5}\right\rfloor,\,0,\,3\right)
\]
from which the change score floor is obtained:
\[
s_{\min}=-3-k
\]
With this design, the resulting range is:
\[
s_{\min} \in [-6,\,-3]
\]
Stage~2 transitions to Stage~3 when $s^{(t)}\ge 3$. In Stage~2 and Stage~3, early termination happens when $s^{(t)}\le s_{\min}$.
\end{multicols}

%% file: architecture_modules_multicolumn.tex
\section{\Name Architecture Details}
\label{app:architecture_modules_multicolumn}
\label{app:architecture_modules}

\begin{multicols}{2}
This appendix includes the functionalities and implementation details of each internal modules of \Name, including the classifier, controllers, and the prompt builder.

\subsection{Classifier Modules}

All three classifiers are implemented as prompt based LLM modules with the same model and inference setup. In code, each classifier call uses OpenAI \texttt{gpt-5.2-2025-12-11} \citep{OpenAIGPT52Docs} with default inference parameters, and each output is constrained to a predefined JSON schema. Full prompts for all classifiers are documented in Appendix~\ref{app:classifier_prompts_multicolumn}.

\paragraph{Counsellor Intent Classifier.}
This classifier reads the latest counsellor utterance and returns six outputs: four Boolean intent flags and two category lists. The four intent flags are whether the counsellor is evoking motivations for change, evoking barriers against change, asking about next steps, or eliciting commitment to specific next steps. The two category lists are for specific DARNCATs labels for the motivations and barriers asked by the counsellor. 

\paragraph{Counsellor Behaviour Classifier.}\label{par:counsellor_behaviour_classifier_multicolumn}
\label{par:counsellor_behaviour_classifier}
This classifier operates at the turn level and labels the latest counsellor turn with MITI behaviour codes \citep{MITI}. For each counsellor turn, it returns an ordered sequence of unique labels drawn from the MITI coding scheme, where labels follow the order in which behaviours first occur in that turn. If no notable behaviour is detected, the output defaults to NC. Please refer to Appendix~\ref{app:classifier_prompts_multicolumn} for the full MITI coding scheme. Note that the "Confront" and "Persuade" counsellor behaviours are the only two that are not adherent to MI principles. If these are detected, it will trigger an override (of the rules described above) in the Reveal Controller.

\paragraph{Client Behaviour Classifier.}
This classifier assigns exactly one DARNCATs (together with + or - as described above) label to each client turn. These labels also correspond to the labels used for motivations and barriers in structured profiles, so that turn-level client signals and profile items are aligned under the same category system.

\subsection{Conversation State Controller}
The Conversation State Controller is a deterministic rule-based algorithm that implements the stage progression and early termination logic defined in Section~\ref{subsec:stage_progression}. The conversation state is a shared variable that encapsulates information required to keep track of and direct stage progression, such as the current stage and speaker, revealed profile items and flags for stage transitions. At each turn, it takes the latest outputs from the Counsellor Intent Classifier and the Client Behaviour Classifier, together with the previously stored conversation state, and updates the conversation state. The same state object is also accessible by the Reveal Controller as a way to communicate and share information. The updated state is passed to the Prompt Builder as control signals for the next client turn. Full algorithmic details are provided in Appendix~\ref{app:state_control_multicolumn}.

\subsection{Reveal Controller}
The Reveal Controller is a deterministic rule-based algorithm that implements the disclosure rules defined in Section~\ref{subsec:reveal_policy}. At each turn, it reads the shared conversation state together with the latest outputs from the Counsellor Intent Classifier and Counsellor Behaviour Classifier, then evaluates stage-gated reveal eligibility, refusal conditions, and counsellor behavior-based overrides. It outputs one control decision for the turn (reveal at most one item, or refuse), writes reveal bookkeeping and refusal updates back to the same shared state object, and emits the control message consumed by the Prompt Builder. Full algorithmic details are provided in Appendix~\ref{app:reveal_control_multicolumn}.

\subsection{Prompt Builder}
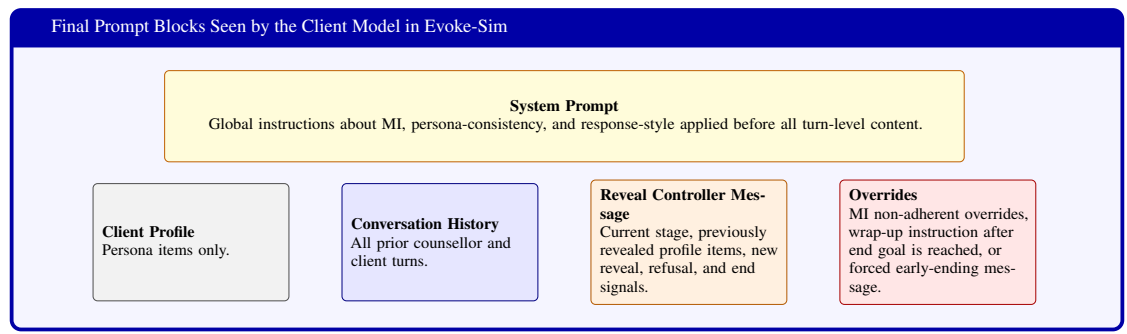
\begin{figure*}[!t]
\centering
\scriptsize
\begin{tcolorbox}[colback=blue!4,colframe=blue!65!black,title={Final Prompt Blocks Seen by the Client Model in \Name},width=0.92\textwidth,fontupper=\small]
\centering
\resizebox{0.92\linewidth}{!}{%
\begin{tikzpicture}[
>=Latex,
block/.style={draw,rounded corners=2pt,align=left,text width=0.20\textwidth,minimum height=18mm,inner sep=4pt,font=\scriptsize},
sysblock/.style={draw,rounded corners=2pt,align=center,text width=0.88\textwidth,minimum height=14mm,inner sep=4pt,font=\scriptsize}
]
\coordinate (rowmid) at (0,0);
\node[block,fill=blue!10,draw=blue!50!black,anchor=east] (b2) at ($(rowmid)+(-4mm,0)$) {\textbf{Conversation History}\\All prior counsellor and client turns.};
\node[block,fill=orange!12,draw=orange!70!black,anchor=west] (b3) at ($(rowmid)+(4mm,0)$) {\textbf{Reveal Controller Message}\\Current stage, previously revealed profile items, new reveal, refusal, and end signals.};
\node[block,fill=gray!10,draw=gray!60!black,left=8mm of b2] (b1) {\textbf{Client Profile}\\Persona items only.};
\node[block,fill=red!10,draw=red!65!black,right=8mm of b3] (b4) {\textbf{Overrides}\\MI non-adherent overrides, wrap-up instruction after end goal is reached, or forced early-ending message.};
\node[sysblock,fill=yellow!18,draw=orange!75!black] (sys) at ($(b1.north)!0.5!(b4.north)+(0,10mm)$) {\textbf{System Prompt}\\Global instructions about MI, persona-consistency, and response-style applied before all turn-level content.};
\end{tikzpicture}
}
\end{tcolorbox}
\caption{Prompt composition for each client turn in \Name.}
\label{fig:evokesim_prompt_blocks_multicolumn}
\label{fig:evokesim_prompt_blocks}
\end{figure*}

The Prompt Builder is the final internal component before client utterance generation. As shown in Figure~\ref{fig:evokesim_prompt_blocks_multicolumn}, each client turn is conditioned by a persistent system prompt plus four functional blocks: client profile, conversation history, reveal controller message, and overrides. In implementation, the system prompt carries global role instructions and persona-consistency constraints, the client profile block contains only persona items, the conversation history block contains all prior counsellor and client turns, and the reveal controller message carries stage, visible items, newly revealed item (if any), and refusal or end signals. The override block is used only when special control conditions are active, which are MI non-adherent behaviour handling, wrap-up instructions after end goal is reached, and forced early-ending turns, through which direct instructions are injected into the prompt. To keep each reply concise and relevant, the prompt enforces single-sentence outputs. Full prompt templates and block-level variants used by this module are provided in Appendix~\ref{app:prompt_builder_prompts_multicolumn}.
\end{multicols}

%% file: classifier_prompts_multicolumn.tex
\section{Classifier Prompt Templates}
\label{app:classifier_prompts_multicolumn}
\label{app:classifier_prompts}

\begin{multicols}{2}
This appendix provides the full prompts used by the three classifier modules described in Section~\ref{subsec:implementation_flow}. All three classifier calls use OpenAI \texttt{gpt-5.2-2025-12-11} \citep{OpenAIGPT52Docs} with default inference settings and schema constrained JSON outputs. Table~\ref{tab:miti_labels_full_multicolumn} summarizes the full MITI label set used by the counsellor behaviour classifier.

\end{multicols}
\nolinenumbers
\begin{table}[H]
\centering
\small
\begingroup
\renewcommand{\arraystretch}{1.50}
\begin{tabular}{p{0.22\linewidth}p{0.72\linewidth}}
\thickhline
\textbf{MITI Label} & \textbf{One-Sentence Definition} \\
\hline
Giving Information (GI) & The counsellor provides neutral information, feedback, or professional opinion without explicitly directing the client towards change. \\
Persuade & The counsellor explicitly tries to influence client attitudes or behavior using arguments, advice, recommendations, or solutions without clear autonomy support. \\
Persuade with Permission (Persuade with) & The counsellor gives directive guidance while also seeking permission or clearly emphasizing client autonomy and collaboration. \\
Question (Q) & The counsellor asks a question unless the utterance is better captured by another MITI behavior code. \\
Simple Reflection (SR) & The counsellor restates or lightly paraphrases the client's explicit content without adding deeper meaning or emotional intensity. \\
Complex Reflection (CR) & The counsellor reflects beyond surface content by adding implied meaning, stronger emotion, integration, or reframing. \\
Affirm (AF) & The counsellor genuinely highlights a specific client strength, effort, intention, value, or accomplishment. \\
Seeking Collaboration (Seek) & The counsellor explicitly shares power by inviting collaboration or agreement on goals, plans, or session direction. \\
Emphasizing Autonomy (Emphasize) & The counsellor explicitly reinforces that decisions and actions belong to the client and affirms freedom of choice. \\
Confront & The counsellor directly disagrees, warns, criticizes, shames, blames, or otherwise challenges the client in a non-collaborative way. \\
No Code (NC) & No MITI behavior code applies, typically for structuring, greetings, facilitative backchannels, incomplete thoughts, or off-topic material. \\
\thickhline
\end{tabular}
\endgroup
\caption{Full MITI behavior label set used by the counsellor behaviour classifier.}
\label{tab:miti_labels_full_multicolumn}
\label{tab:miti_labels_full}
\end{table}
\makeatletter
\@ifundefined{promptbox}{
\newtcolorbox{promptbox}[1]{
    breakable,
    colback=blue!4,
    colframe=blue!65!black,
    width=\linewidth,
    fontupper=\small,
    title={#1}
}
}{}
\makeatother

\lstdefinestyle{appendixprompt}{
  basicstyle=\small\rmfamily,
  breaklines=true,
  breakindent=0pt,
  columns=fullflexible,
  keepspaces=true,
  showstringspaces=false
}

\begin{promptbox}{Box D.1: Full Prompt for the Counsellor Intent Classifier}
\begin{lstlisting}[style=appendixprompt]
You are a classifier for motivational interviewing (MI) counsellor intent.

You will be given a short transcript of a conversation between a counsellor and a client.
The counsellor is trying to help the client change their smoking behavior.
In the latest counsellor utterance, the counsellor may have asked a question, and perhaps it was trying to elicit certain information from the client.
Your task is to classify whether the *QUESTION* in the *latest counsellor utterance* is doing each of the following (each can be true/false independently):

1) eliciting_motivations_for_change:
   - asking about reasons to change, benefits of change, desire/need/importance, values/goals that support change.
     e.g., "What are the benefits of quitting smoking?", "What are some of the downsides of smoking?", "What makes quitting smoking important to you?".

2) eliciting_barriers_against_change:
   - asking about challenges, obstacles, downsides of change, reasons for continuing, ambivalence, triggers, what makes change hard.
     e.g., "What makes quitting smoking difficult for you?", "What made you come back to your old smoking habit?", "What are some of the challenges you face when trying to quit?".

3) asking_next_steps:
   - asking about potential next steps and plans (e.g., "What do you think you will do?", "Where would you like to go from here?", "What are the next steps you had in mind?").

4) eliciting_commitment:
   - asking for explicit commitment/decision (e.g., "would you like to commit to this plan", "are you ready to decide to commit to these next steps", "can you agree to...").

If in the latest counsellor utterance there is no question, or if the question does not pertain to any of the above four intents, mark all four as false.

The counsellor may have also intended to ask the CLIENT about specific motivation (+) or barrier (-) categories. These categories are:
Motivation (+) categories:
- Desire+: the client's willingness to alter the smoking behavior.
- Ability+: the client's ability to change the smoking behavior.
- Reason+: the client's rationale for changing the smoking behavior.
- Need+: the client's need to change the smoking behavior.
- Commitment+: the client's commitment to change the smoking behavior.
- Activation+: if the client is leaning towards action about changing the smoking behavior.
- Taking Steps+: the steps the client has already taken in the recent past (within approximately the past week) towards changing the smoking behavior.
- Other+: other motivations for changing the smoking behavior. Includes hypothetical situations or
  circumstances that would convince the client to change, and problem recognition.

Barrier (-) categories:
- Desire-: the client's unwillingness to change or wish to partake in the smoking behavior.
- Ability-: the client's inability or lack of confidence about changing the smoking behavior.
- Reason-: the client's rationale for not changing the smoking behavior or for why change is unnecessary.
- Need-: the client's need not to change the smoking behavior or to stay the same.
- Commitment-: the client's commitment/decision not to change the smoking behavior, or an idea for how not to change/to stay the same.
- Activation-: if the client is leaning towards not taking any action about changing the smoking behavior.
- Taking Steps-: the steps the client has already taken in the recent past (within approximately the past week) resisting changing the smoking behavior.
- Other-: Other barriers against changing the smoking behavior. This includes minimization of problems and hypothetical statements about non-change.

If the counsellor explicitly asks the CLIENT for any of the above specific motivation (+) or barrier (-) categories, list them:
- requested_plus_categories: any of [Desire+, Ability+, Reason+, Need+, Commitment+, Activation+, TakingSteps+, Other+]
- requested_minus_categories: any of [Desire-, Ability-, Reason-, Need-, Commitment-, Activation-, TakingSteps-, Other-]
If none are specific, return empty lists.

Return STRICT JSON (no markdown) with exactly these keys:
{{
  "eliciting_motivations_for_change": true/false,
  "eliciting_barriers_against_change": true/false,
  "asking_next_steps": true/false,
  "eliciting_commitment": true/false,
  "requested_plus_categories": [...],
  "requested_minus_categories": [...]
}}

Notes:
- Any or all of the flags may be false; it is valid for all four to be false if none of these intents apply to the latest counsellor turn.
- The requested_* lists should only include categories the counsellor explicitly asked the CLIENT to talk about.

Transcript:
{transcript}
\end{lstlisting}
\end{promptbox}

\begin{promptbox}{Box D.2: Full Prompt for the Counsellor Behaviour Classifier}
\begin{lstlisting}[style=appendixprompt]
You are a classifier for Motivational Interviewing (MI) counsellor behavior using MITI-style labels.

You will be given a short transcript of a conversation between a counsellor and a client.
Your task is to classify the *latest counsellor utterance*.
The utterance may contain more than one code. Return all applicable labels in the order they occur in the utterance.

IMPORTANT RULES:
- Output a list of labels with no repeats.
- The labels must be in the order they FIRST occur in the utterance.
- If any other label applies, do NOT include NC.
- If no label applies, return only ["NC"].

Classification Categories:
 - **Giving Information (GI)**: Used when the interviewer gives information, educates, provides feedback, or expresses a professional opinion without persuading, advising, or warning. Typically, the tone of the information is neutral, and the language used to convey general information does not imply that it is specifically relevant to the client or that the client must act on it. NOTE: Structuring statements are not coded as Giving Information. These include statements that indicate what is going to happen during the session, instructions for an exercise during the session, set-up of another appointment, or discussion about the number and timing of sessions for a research protocol. Also, Giving information should not be confused with persuading, confronting, or persuading with permission.

 - **Persuade (Persuade)**: The clinician makes overt attempts to change the client's opinions, attitudes, or behavior using tools such as logic, compelling arguments, self-disclosure, or facts (and the explicit linking of these tools with an overt message to change). Persuasion is also coded if the clinician gives biased information, advice, suggestions, tips, opinions, or solutions to problems without an explicit statement or strong contextual cue emphasizing the client's autonomy in receiving the recommendation. Note that if the therapist is giving information in a neutral manner, without an explicit focus on influencing or convincing the client, the Giving Information code should be used. Decision Rule: If the coder cannot decide between the Persuasion and the Giving Information code, the Giving Information code should be used. This decision rule is intended to set a relatively high bar for the Persuasion code.

 - **Persuade With Permission (Persuade_with)**: Persuade with Permission is assigned when the interviewer includes an emphasis on collaboration or autonomy support while persuading. The condition of permission may be present when 1. The client asks directly for the clinician's opinion on what to do or how to proceed. 2. The clinician asks the client directly for permission to provide advice, make suggestions, give opinion, offer feedback, express concerns, make recommendations, or discuss a particular topic. 3. The clinician uses autonomy-supportive language to preface or qualify the advice such that the client may choose to discount, ignore, or personally evaluate that advice. The clinician could seek a general sense of permission (How about we start today talking about your probation requirements?) or permission specific to a topic, condition, or action item (If it is alright with you, I'll share some strategies that have been used by others to keep their blood sugar in check.). Permission may be obtained before, during, or after persuasion is used, but must occur close to persuasion in time. If Persuade with Permission is accompanied by an explicit Seeking Collaboration or Emphasizing Autonomy, both the Persuade with Permission and the Seeking Collaboration or Emphasizing Autonomy code should be assigned. If a clinician has asked for more general permission, it does not need to be repeated for every statement or suggestion. There is a "condition of permission" that may last for several minutes (or several volleys). If the clinician changes the topic, becomes more directive, starts adding significant content (becomes the expert), or starts prescribing a plan without again asking permission, it is possible that the clinician would then receive a Persuade code. Note that if the interviewer is providing information or advice in a neutral manner, the Giving Information code should be used instead. If the coder is uncertain, the GI code should be preferred. Decision Rule: When both Persuade AND Persuade with Permission occur in the same utterance, the coder should only assign the Persuade with Permission code.To better determine whether Persuade with Permission is a more suitable code compared to Persuade, use the context of a couple of volleys (e.g. 4-5 volleys) before the current volley, as provided from the provided excerpt of a MI counselling session transcript, to determine if permission was sought by the client or counsellor (for example, if the client explicitly asks the counsellor for advice on a topic e.g. "I was wondering if you could tell me"), or if autonomy-supportive language was used, in relation to the persuade utterance. It is key to look at whether the client asked for advice in the previous couple volleys, as this is often missed when coding Persuade with Permission. If the counsellor's advice is given in response to a question seeking advice from the client, this counts as Persuade with Permission.

 - **Question (Q)**: All questions from clinicians (open, closed, evocative, fact-finding, etc.) receive the Question code, except if they belong in any other code (e.g. Seeking Collaboration).

 - **Simple Reflection (SR)**: Simple reflections restate or lightly paraphrase the client's explicit words without adding emotional intensity, implied meaning, new perspective, or motivational significance. SRs track the client's surface message, whether factual or emotional, using similar strength, tone, and intention, and they should not clarify unspoken feelings, highlight deeper themes, or strengthen the emotional language. Even if wording changes, the reflection remains SR only when it mirrors exactly what the client already made explicit, provides no additional insight, and does not guide the client towards deeper awareness. Any reflection that subtly sharpens feelings, interprets meaning, or expands the client's experience should not be labeled SR.

 - **Complex Reflection (CR)**: Complex reflections deepen or extend the client's message by highlighting implied emotions, strengthening emotional language, clarifying underlying meaning or motivation, integrating multiple client ideas, reframing content, or adding interpretive emphasis. CRs go beyond surface paraphrasing and help the client see their experience more clearly or in a more psychologically meaningful way, even when the added meaning is subtle. When the therapist selects stronger emotional words, articulates a feeling the client only hinted at, emphasizes a key theme, or conveys a deeper interpretation that organizes or expands the client's narrative, it should be coded as CR. If a reflection offers any meaningful elaboration beyond what the client explicitly stated, it is CR rather than SR.

 - **Affirm (AF)**: A clinician utterance that accentuates something positive about the client. To be considered an Affirm, the utterance must be about the client's strengths, efforts, intentions, or worth. The utterance must be given in a genuine manner and reflect something genuine about the client. It does not have to be focused on the change goal and could reflect a "prizing" of the client for a specific trait, behavior, accomplishment, skill, or strength. Affirms are often complex reflections, and when this occurs, the Affirm code should be preferred. Affirm should not be coded automatically for the clinician's agreeing with, approval of, cheerleading for, or non-specific praising of the client. They must be explicitly linked to client behaviors or specific characteristics. The utterance must seem genuine and not merely facilitative. If the coder is not certain whether the statement is specific or strong enough to merit the Affirm code, it should not be assigned.

 - **Seeking Collaboration (Seek)**: This code is assigned when a clinician explicitly attempts to share power or acknowledge the expertise of the client. It can occur when the clinician genuinely seeks consensus with the client regarding tasks, goals, or directions of the session. Seeking Collaboration may be assigned when the clinician asks what the client thinks about information provided. When permission to give information or advice is sought, Seeking Collaboration is typically assigned. When a clinician asks about the client's knowledge or understanding of a particular topic, this is coded as a Question. It is not considered to be Seeking Collaboration.

 - **Emphasizing Autonomy (Emphasize)**: These are utterances that clearly focus the responsibility with the client for decisions about and actions pertaining to change. They highlight clients' sense of control, freedom of choice, personal autonomy, or ability or obligation to decide about their attitudes and actions. These are not statements that specifically emphasize the client's sense of self-efficacy, confidence, or ability to perform a specific action.

 - **Confront (Confront)**: This code is used when the clinician confronts the client by directly and unambiguously disagreeing, arguing, correcting, shaming, blaming, criticizing, labeling, warning, moralizing, ridiculing, or questioning the client's honesty. Such interactions will have the quality of uneven power sharing, accompanied by disapproval or negativity. Included here are instances where the interviewer uses a question or even a reflection, but the voice tone clearly indicates a confrontation. Restating negative information already known or disclosed by the client can be either a Confront or a Reflection. Most Confronts can be correctly categorized by careful attention to voice tone and context. When clinicians use confrontation to emphasize a client strength, virtue, or positive achievement, the Affirm code should be considered. A Confront is not mandatory when the clinician is clearly attempting to affirm or support the client.
 
 - **No Code (NC)**: When none of the above codes applies. The MITI is not an exhaustive coding system because some utterances may not receive a behavior code. Examples of utterances that are not coded in the MITI include: structure statements ("Now we'll talk about the forms from last week"), greetings ("Hi Joe. Thanks for coming in today"), facilitative statements ("Okay, all right. Good"), references to previous session content ("Last week you mentioned you were really tired"), incomplete thoughts ("You mentioned..." [client interrupts]), and off-topic material ("It's a bit cold in here").

Acceptable label outputs are strictly within this set:
["GI", "Persuade", "Persuade_with", "Q", "SR", "CR", "AF", "Seek", "Emphasize", "Confront", "NC"]

Return STRICT JSON (no markdown) with exactly one key:
{{
  "labels": ["..."]
}}

Transcript:
{transcript}
\end{lstlisting}
\end{promptbox}

\begin{promptbox}{Box D.3: Full Prompt for the Client Behaviour Classifier}
\begin{lstlisting}[style=appendixprompt]
You are a classifier for Motivational Interviewing (MI) client language.

## DARNCAT Definitions and Examples (adapted to smoking)

Neutral client language. Neutral, or non-change, client language gets the Neutral code. It must be distinguished from change talk (CT) and counter-change talk (CCT). Neutral includes:
- Questions asked of the therapist that are not themselves change/counter-change statements:
  - "What do you think I should do about smoking?"
- Reporting of factual information (e.g. number of cigarettes per week):
  - "Sometimes on Fridays I'll have a few cigarettes with friends."
- Story-telling unrelated to current change in quitting smoking:
  - "I was downtown with my partner and we ran into friends for dinner; we talked, I had one cigarette and went home early."
- Behaviors/events occurring in the distant past (defined as more than approximately a week prior to the current therapy session):
  - "In high school I tried quitting for a month after getting sick."
- Talking about someone else's intentions to change/not change:
  - "My brother is thinking about quitting, and I think he needs to. He smokes way too much."
- Language that indicates the client is following the therapist but does not indicate agreement with the therapist:
  - "Uh huh."
  - "OK."
- Any other client language that is neither CT nor CCT
  - "I'm going to need to leave a little early today because my daughter has soccer practice."
  - "I'd like a tissue."

Counter-change talk. This type of client language refers to any movement away from change, or towards
sustaining the target behavior. Note that "change" here is defined in reference to the target behavior.
Within the context of treatment for problem smoking, for example, counter-change talk is coded in
relation to maintaining or increasing smoking behavior. Clients may express counter-change talk on
other subjects (e.g., change in a relationship, moving to a new apartment), but these are not coded
unless directly related to the identified target behavior change. Counter-change talk need not have an
oppositional quality nor an emotional charge. The key is that the client language favors not changing
the target behavior, representing status quo or movement backward. Endorsing or expressing agreement
with counter-change talk offered by the therapist should be coded as an instance of counter-change
talk.
Each different counter-change talk statement counts as one instance of counter-change talk. For
example, if a client lists several different reasons against or disadvantages of change, each one is
coded as counter-change talk (e.g., a volley that included a Desire-, Need-, and Other- would count as
three counter-change talk tallies, and a string of four Reason-'s would count as four counter-change
talk tallies).
Some sub-categories of counter-change talk include:

- Reason-: A statement indicating a rationale for not changing or for why change is unnecessary.
  - "Going out with friends wouldn't be any fun without having a smoke together."
  - "The kids stress me out too much when I'm not smoking."
  - "My health seems fine, so I don't see a problem with my smoking."
- Desire-: A special type of reason, expressing the client's unwillingness to change or wish to
  partake in the target behavior.
  - "If I could, I would smoke every day until I'm 90."
  - "I love smoking."
- Need-: A special type of reason stating a need not to change or to stay the same.
  - "Treatment for smoking isn't something that I need right now."
  - "I don't need to quit smoking entirely."
  - "I need to keep smoking if I want to keep these friends."
- Ability-: A statement that client is unable or unconfident about change.
  - "It's just too hard to change my smoking after so many years."
  - "I'm feeling pretty low on the confidence scale about quitting."
- Commitment-: A statement that the client will not change, or an idea for how not to change/to stay
  the same.
  - "As soon as I get out of this program I'm going to buy another pack."
  - "I'm not going to say that I won't smoke ever again."
- Activation-: A statement that the client is leaning towards not making changes or not open to change, 
  but has not committed to not changing just yet.
  - "I'm not willing to try what you just suggested"
  - "I would not consider quitting at least for this month."
  - "I'm not really thinking about smoking less."
- Taking Steps-: A statement that the client is already resisting change; this represents steps taken
  in the recent past (within approximately the past week).
  - "I picked up another carton of cigarettes over the weekend."
  - "I stopped going to my smoke-free support group."
- Other-: A statement that is clearly counter-change talk but does not fit reasonably into the other
  categories. This includes minimization of problems and hypothetical statements about non-change.
  - "Another warning from my doctor about smoking isn't that big of a deal to me."
  - "If I were allowed to smoke at work, I'd light up right at my desk."

Change talk. This type of client language refers to any movement towards change or away from the target
behavior. As with counter-change talk, "change" here is defined specifically in reference to the target
behavior. The client makes a statement that directly or indirectly shows evidence of at least one of
the following categories, which have the quality of moving forward in the direction of change in the
target behavior. Within the context of treatment for problem smoking, for example, change talk is coded
in reference to reducing or stopping smoking behavior.
Each different change talk statement counts as one instance of change talk. For example, if a client
lists several different reasons for or advantages of change, each one is coded as change talk. As with
counter-change talk, endorsing or expressing agreement with change talk offered by the therapist should
be coded as an instance of change talk.
Some sub-categories of change talk include:

- Reason+: A statement indicating a rationale for changing the target behavior.
  - "Quitting smoking would help me feel less distracted at work."
  - "I hate the way my lungs feel after I smoke."
  - "My family needs me to be present at home, not outside smoking."
- Desire+: A special type of reason stating the client's willingness to alter the target behavior.
  - "I really want to get started with quitting smoking."
  - "I don't even feel like having a cigarette today."
- Need+: A special type of reason stating the client's need to change.
  - "I have to do this and quit smoking."
  - "Getting help to stop smoking is what I need right now."
- Ability+: A statement indicating that the client is able to change.
  - "I know that I can quit if I try hard enough."
  - "This doesn't seem so difficult if I take it one day at a time."
- Commitment+: A statement that the client will change, or an idea for how the client could change.
  - "I'll do whatever it takes to cut down on my smoking."
  - "I could start by tossing out all the cigarettes at home."
- Activation+: A statement that the client is leaning towards action and open to change, but has not yet
  made a clear commitment.
  - "I'm willing to give it another try."
  - "I would consider quitting this month."
  - "I'm thinking about reducing the amount I smoke."
- Taking Steps+: A statement that the client has already begun to change; this represents steps taken
  in the recent past (within approximately the past week).
  - "At dinner last night I told my family that I'm going to quit smoking."
  - "I've already cut down the number of cigarettes I smoke this week."
- Other+: Any other statement about changing the target behavior. Includes hypothetical situations or
  circumstances that would convince the client to change, and problem recognition.
  - "My smoking is out of control."
  - "If I could move to a smoke-free environment, I'd be less likely to feel the urge to smoke."

## DARNCAT Label Set for Coding
When you classify motivations for change or barriers against change, use ONLY the following category
labels:
- Desire+, Ability+, Reason+, Need+, Commitment+, Activation+, TakingSteps+, Other+
- Desire-, Ability-, Reason-, Need-, Commitment-, Activation-, TakingSteps-, Other-

Base all categorizations only on what the client actually says about smoking.

You will be given a short transcript about changing the client's smoking behavior. Your task is to classify the *latest client utterance*.

Return STRICT JSON (no markdown) with exactly one key:
{{
  "label": "<one of: Neutral, Desire+, Ability+, Reason+, Need+, Commitment+, Activation+, TakingSteps+, Other+, Desire-, Ability-, Reason-, Need-, Commitment-, Activation-, TakingSteps-, Other->"
}}

Rules:
- Output exactly ONE label.
- Use '+' labels ONLY if the utterance clearly moves towards changing smoking (reducing/stopping).
- Use '-' labels ONLY if the utterance supports continuing smoking / resisting change (sustain/counter-change talk).
- Use Neutral if the utterance is neither change talk nor counter-change talk.

Transcript:
{transcript}
\end{lstlisting}
\end{promptbox}

%% file: state_control_multicolumn.tex
\section{Conversation State Control Algorithm}
\label{app:state_control_multicolumn}
\label{app:state_control}

\begin{multicols}{2}
This appendix specifies the exact deterministic conversation state control algorithm used by \Name. The Conversation State Controller and Reveal Controller read from and write to the same runtime state object, so all conversation state variables are updated jointly across turns. Refusal count and revealed item sets are written by the Reveal Controller to the shared runtime state, and the state controller reads them to derive pending end signals. Table~\ref{tab:state_control_io_multicolumn} summarizes the symbols used in this appendix. Algorithm~\ref{alg:state_control_update_multicolumn} describes how the conversation state controller takes in the state variables through $S^{(t-1)}$, updates shared state to $S^{(t)}$, and emits control payload $O^{(t)}$ for downstream modules at the end of the turn.

\end{multicols}
\nolinenumbers

\begin{table}[H]
\centering
\small
\begingroup
\renewcommand{\arraystretch}{1.25}
\begin{tabular}{p{0.33\linewidth}p{0.62\linewidth}}
\thickhline
\textbf{Symbol} & \textbf{Definition} \\
\hline
\multicolumn{2}{l}{\textit{Input Signals and Constants}} \\
$S^{(t-1)}$ & Conversation state before the update at turn $t$. \\
$I^{(t)}$ & Counsellor intent output at turn $t$ (motivation request, barrier request, next step request, commitment request, and requested DARNCATs categories). \\
$y^{(t)}$ & Client behaviour label at turn $t$ (one DARNCATs label). \\
$C_{+}, C_{-}$ & Plus and minus DARNCATs label sets, following Table~\ref{tab:darncats_labels_multicolumn} in Appendix~\ref{app:profile_extraction_multicolumn}. \\
$s_{\min}$ & Change score floor for the profile, defined in Appendix~\ref{app:change_score_multicolumn}. \\
\hline
\multicolumn{2}{l}{\textit{Shared State Fields}} \\
$S^{(t)}$ & Conversation state after the update at turn $t$. \\
$S^{(t)}.\mathrm{speaker}$ & Speaker field in state at turn $t$. \\
$S^{(t)}.\mathrm{stage}$ & Stage field in state at turn $t$ ($1$, $2$, or $3$). \\
$S^{(t)}.\mathrm{revealedNextSteps}$ & Revealed next-step items currently stored in state. \\
$S^{(t)}.\mathrm{Stage1ClientTurnCount}$ & Number of client turns processed while the conversation is in Stage~1. \\
$S^{(t)}.\mathrm{SeenPlusInStage1}$ & Stage~1 flag that indicates whether a plus client label has been observed. \\
$S^{(t)}.\mathrm{SeenMinusInStage1}$ & Stage~1 flag that indicates whether a minus client label has been observed. \\
$S^{(t)}.\mathrm{RefusalCount}$ & Number of end-eligible refusals accumulated so far in the conversation (stage-gated next-step or commitment refusals in Stages~1 and~2). \\
\hline
\multicolumn{2}{l}{\textit{Derived Variables}} \\
$W_{\mathrm{C}}^{(t-1)}, W_{\mathrm{C}}^{(t)}$ & Stage 2/3 change score window before and after update (up to 10 recent client labels). \\
$\#C_{+}^{(t)},\#C_{-}^{(t)}$ & Counts of labels in $W_{\mathrm{C}}^{(t)}$ that belong to $C_{+}$ and $C_{-}$, respectively, as defined in Appendix~\ref{app:change_score_multicolumn}. \\
$s^{(t)}$ & Change score at turn $t$, defined in Appendix~\ref{app:change_score_multicolumn}. \\
$g^{(t)}$ & End goal flag at turn $t$. \\
\hline
\multicolumn{2}{l}{\textit{Output}} \\
$O^{(t)}$ & Controller output payload for downstream modules; it contains the current stage, change score, change and sustain counts, end goal flag, and runtime control bookkeeping from state (pending end signals, refusal and MI non-adherent counts, reveal counts, and visible profile IDs). \\
\thickhline
\end{tabular}
\endgroup
\caption{Inputs, outputs, and intermediate variables for conversation state control.}
\label{tab:state_control_io_multicolumn}
\label{tab:state_control_io}
\end{table}

\newcounter{appendixalgorithm}
\renewcommand{\theappendixalgorithm}{\thesection.\arabic{appendixalgorithm}}
\renewcommand{\theHappendixalgorithm}{\thesection.\arabic{appendixalgorithm}}
\refstepcounter{appendixalgorithm}\label{alg:state_control_update_multicolumn}
\label{alg:state_control_update}
\begin{table}[H]
\centering
\small
\begingroup
\renewcommand{\arraystretch}{1.18}
\setlength{\tabcolsep}{4pt}
\begin{tabular}{p{0.045\linewidth}p{0.92\linewidth}}
\thickhline
\multicolumn{2}{l}{\textbf{Algorithm~\theappendixalgorithm\; Conversation State Control Update (per turn)}} \\
\thickhline
\multicolumn{2}{p{0.965\linewidth}}{\textbf{Inputs.} $S^{(t-1)}$, $I^{(t)}$, $y^{(t)}$, and precomputed profile floor $s_{\min}$.} \\
\multicolumn{2}{p{0.965\linewidth}}{\textbf{Outputs.} Updated state $S^{(t)}$ and control payload $O^{(t)}$ for Reveal Controller and Prompt Builder.} \\
\hline
1 & \textit{/* Update counsellor side signals */} \\
2 & $S^{(t)} \leftarrow S^{(t-1)}$ \\
3 & \textbf{if } $S^{(t)}.\mathrm{speaker}=\text{counsellor}$ \\
4 & \hspace*{1.5em}$S^{(t)} \leftarrow \mathrm{UpdateIntentState}(S^{(t)}, I^{(t)})$ \\
& \\
5 & \textit{/* Update client side tracking */} \\
6 & \textbf{if } $\mathrm{HasClientLabel}(y^{(t)})$ \\
7 & \hspace*{1.5em}\textbf{if } $S^{(t)}.\mathrm{stage}=1$ \\
8 & \hspace*{3.0em}$S^{(t)}.\mathrm{Stage1ClientTurnCount} \mathrel{+}= 1$ \\
9 & \hspace*{3.0em}$S^{(t)} \leftarrow \mathrm{UpdateStage1AmbivalenceFlags}(S^{(t)}, y^{(t)}, C_{+}, C_{-})$ \\
10 & \hspace*{1.5em}\textbf{else} \\
11 & \hspace*{3.0em}$W_{\mathrm{C}}^{(t)} \leftarrow \mathrm{UpdateChangeWindow}(W_{\mathrm{C}}^{(t-1)}, y^{(t)}, 10)$ \\
12 & \hspace*{3.0em}$\#C_{+}^{(t)}, \#C_{-}^{(t)} \leftarrow \mathrm{CountClientLabels}(W_{\mathrm{C}}^{(t)}, C_{+}, C_{-})$ \\
13 & \hspace*{3.0em}$s^{(t)} \leftarrow \mathrm{ComputeChangeScore}(\#C_{+}^{(t)}, \#C_{-}^{(t)}, s_{\min})$ \\
14 & \hspace*{3.0em}$S^{(t)} \leftarrow \mathrm{UpdateScoreState}(S^{(t)}, W_{\mathrm{C}}^{(t)}, \#C_{+}^{(t)}, \#C_{-}^{(t)}, s^{(t)})$ \\
& \\
15 & \textit{/* Update end goal signal */} \\
16 & \textbf{if } $S^{(t)}.\mathrm{stage}=3 \land \mathrm{IsCommitmentPlus}(y^{(t)})$ \\
17 & \hspace*{1.5em}$g^{(t)} \leftarrow \mathrm{CheckEndGoal}(S^{(t)}.\mathrm{revealedNextSteps})$ \\
18 & \hspace*{1.5em}$S^{(t)} \leftarrow \mathrm{UpdateEndGoalState}(S^{(t)}, g^{(t)})$ \\
& \\
19 & \textit{/* Apply stage transitions */} \\
20 & $s^{(t)} \leftarrow \mathrm{ReadChangeScore}(S^{(t)})$ \\
21 & \textbf{if } $S^{(t)}.\mathrm{stage}=1 \land S^{(t)}.\mathrm{SeenPlusInStage1} \land S^{(t)}.\mathrm{SeenMinusInStage1}$ \\
22 & \hspace*{1.5em}$S^{(t)}.\mathrm{stage} \leftarrow 2$ \\
23 & \textbf{else if } $S^{(t)}.\mathrm{stage}=1 \land S^{(t)}.\mathrm{Stage1ClientTurnCount}\ge 10$ \\
24 & \hspace*{1.5em}$S^{(t)} \leftarrow \mathrm{SetNoAmbivalenceEndFlag}(S^{(t)})$ \\
25 & \textbf{if } $S^{(t)}.\mathrm{stage}=2 \land s^{(t)} \ge 3$ \\
26 & \hspace*{1.5em}$S^{(t)}.\mathrm{stage} \leftarrow 3$ \\
27 & \textbf{if } $S^{(t)}.\mathrm{stage}\in\{1,2\} \land S^{(t)}.\mathrm{RefusalCount}\ge 3$ \\
28 & \hspace*{1.5em}$S^{(t)} \leftarrow \mathrm{SetRefusalPendingEndFlag}(S^{(t)})$ \\
29 & \textbf{if } $S^{(t)}.\mathrm{stage}\ge 2 \land s^{(t)} \le s_{\min}$ \\
30 & \hspace*{1.5em}$S^{(t)} \leftarrow \mathrm{SetSustainDominanceEndFlag}(S^{(t)})$ \\
& \\
31 & \textit{/* Emit controller outputs */} \\
32 & $O^{(t)} \leftarrow \mathrm{BuildOutputPayload}(S^{(t)})$ \\
33 & $\textbf{return } (S^{(t)}, O^{(t)})$ \\
\thickhline
\end{tabular}
\endgroup
\caption*{Exact rule-based conversation state update and output emission algorithm used by \Name.}
\end{table}

%% file: reveal_control_multicolumn.tex
\section{Reveal Control Algorithm}
\label{app:reveal_control_multicolumn}
\label{app:reveal_control}

\begin{multicols}{2}
This appendix specifies the exact deterministic reveal control algorithm used by \Name. The Reveal Controller and Conversation State Controller read from and write to the same runtime state object, so reveal bookkeeping and stage bookkeeping stay synchronized at each turn. Table~\ref{tab:reveal_control_io_multicolumn} summarizes the symbols used in this appendix. Algorithm~\ref{alg:reveal_control_update_multicolumn} describes the per-turn reveal update that takes in $S^{(t-1)}$ and emits updated $S^{(t)}$ together with the reveal decision for that turn. We separate any turn-level refusal from end-eligible refusals: only refusals for next steps or commitment requests in Stages~1 and~2 increment the shared refusal counter used by early termination.

\end{multicols}
\nolinenumbers

\begin{table}[H]
\centering
\small
\begingroup
\renewcommand{\arraystretch}{1.25}
\begin{tabular}{p{0.33\linewidth}p{0.62\linewidth}}
\thickhline
\textbf{Symbol} & \textbf{Definition} \\
\hline
\multicolumn{2}{l}{\textit{Input Signals and Profile}} \\
$S^{(t-1)}$ & Shared conversation state before reveal update at turn $t$. \\
$I^{(t)}$ & Counsellor intent output at turn $t$ (intent flags and requested DARNCATs categories). \\
$H^{(t)}$ & Counsellor behaviour output at turn $t$ (MITI label sequence). \\
$\Pi$ & Structured profile corresponding to the current client. \\
\hline
\multicolumn{2}{l}{\textit{Shared State Fields}} \\
$S^{(t)}$ & Shared conversation state after reveal update at turn $t$. \\
$S^{(t)}.\mathrm{stage}$ & Current conversation stage in shared state ($1$, $2$, or $3$). \\
$S^{(t)}.\mathrm{MotivationIntentCount}$ & Cumulative motivation-intent count stored in shared state. \\
$S^{(t)}.\mathrm{BarrierIntentCount}$ & Cumulative barrier-intent count stored in shared state. \\
\hline
\multicolumn{2}{l}{\textit{Derived Variables}} \\
$M^{(t)}, B^{(t)}, N^{(t)}, K^{(t)}$ & Motivation, barrier, next-step, and commitment request flags parsed from $I^{(t)}$. \\
$Q_{+}^{(t)}, Q_{-}^{(t)}$ & Requested plus and minus DARNCATs categories parsed from $I^{(t)}$. \\
$n_{\mathrm{req}}^{(t)}$ & Number of requested items implied by the turn-level intent output at turn $t$. \\
$v^{(t)}$ & Behaviour-override flag (true when counsellor behaviour triggers MITI override). \\
$Q_{+, \mathrm{prep}}^{(t)}, Q_{-, \mathrm{prep}}^{(t)}$ & Requested plus and minus categories after preparatory-only filtering in Stages~1 and~2. \\
$\tilde{\Pi}_{M}^{(t)}$ & Unrevealed motivation candidates allowed at turn $t$ after stage filtering. \\
$\tilde{\Pi}_{B}^{(t)}$ & Unrevealed barrier candidates allowed at turn $t$ after stage filtering. \\
$\tilde{\Pi}_{N}^{(t)}$ & Unrevealed next-step candidates allowed at turn $t$. \\
$\rho^{(t)}$ & Requested reveal type selected for this turn (motivation, barrier, or next step). \\
\hline
\multicolumn{2}{l}{\textit{Decision and Outputs}} \\
$r^{(t)}$ & Newly revealed item descriptor for this turn, or $\emptyset$ if no reveal happens. \\
$f^{(t)}$ & Turn-level refusal decision flag for this turn (any refusal type). \\
$f_{\mathrm{end}}^{(t)}$ & End-eligible refusal flag for this turn (only stage-gated next-step or commitment refusal in Stages~1 and~2). \\
$F^{(t)}$ & Combined refusal output for this turn, with $F^{(t)}=(f^{(t)}, f_{\mathrm{end}}^{(t)})$. \\
\thickhline
\end{tabular}
\endgroup
\caption{Inputs, outputs, and intermediate variables for reveal control.}
\label{tab:reveal_control_io_multicolumn}
\label{tab:reveal_control_io}
\end{table}

\setcounter{appendixalgorithm}{0}
\renewcommand{\theappendixalgorithm}{\thesection.\arabic{appendixalgorithm}}
\renewcommand{\theHappendixalgorithm}{\thesection.\arabic{appendixalgorithm}}
\refstepcounter{appendixalgorithm}\label{alg:reveal_control_update_multicolumn}
\label{alg:reveal_control_update}
\begin{table}[H]
\centering
\small
\begingroup
\renewcommand{\arraystretch}{1.14}
\setlength{\tabcolsep}{4pt}
\begin{tabular}{p{0.045\linewidth}p{0.92\linewidth}}
\thickhline
\multicolumn{2}{l}{\textbf{Algorithm~\theappendixalgorithm\; Reveal Control Update (per turn)}} \\
\thickhline
\multicolumn{2}{p{0.965\linewidth}}{\textbf{Inputs.} $S^{(t-1)}$, $I^{(t)}$, $H^{(t)}$, and profile $\Pi$.} \\
\multicolumn{2}{p{0.965\linewidth}}{\textbf{Outputs.} Updated shared state $S^{(t)}$ and reveal decision tuple $(F^{(t)}, r^{(t)})$.} \\
\hline
1 & \textit{/* Read signals and initialize decisions */} \\
2 & $S^{(t)} \leftarrow S^{(t-1)}$ \\
3 & $M^{(t)}, B^{(t)}, N^{(t)}, K^{(t)}, Q_{+}^{(t)}, Q_{-}^{(t)} \leftarrow \mathrm{ReadIntent}(I^{(t)})$ \\
4 & $n_{\mathrm{req}}^{(t)} \leftarrow \mathrm{CountRequestedItems}(I^{(t)})$ \\
5 & $v^{(t)} \leftarrow \mathrm{IsMINonAdherentOverride}(H^{(t)})$ \\
6 & $f^{(t)} \leftarrow \mathrm{false},\;f_{\mathrm{end}}^{(t)} \leftarrow \mathrm{false},\;r^{(t)} \leftarrow \emptyset$ \\
7 & $\tilde{\Pi}_{M}^{(t)},\tilde{\Pi}_{B}^{(t)},\tilde{\Pi}_{N}^{(t)} \leftarrow \mathrm{InitializeCandidatePools}(\Pi, S^{(t)})$ \\
& \\
8 & \textit{/* Refuse broad requests that ask for too many items in one turn */} \\
9 & \textbf{if } $n_{\mathrm{req}}^{(t)} > 2 \land \neg v^{(t)}$ \\
10 & \hspace*{1.5em}$f^{(t)} \leftarrow \mathrm{true}$ \\
& \\
11 & \textit{/* Apply stage-gated refusal for next-steps and commitment in Stages 1-2 */} \\
12 & \textbf{if } $S^{(t)}.\mathrm{stage} \in \{1,2\} \land (N^{(t)} \lor K^{(t)}) \land \neg v^{(t)}$ \\
13 & \hspace*{1.5em}$f^{(t)} \leftarrow \mathrm{true},\;f_{\mathrm{end}}^{(t)} \leftarrow \mathrm{true}$ \\
& \\
14 & \textit{/* Enforce preparatory-category gate in Stages 1 and 2 */} \\
15 & \textbf{if } $S^{(t)}.\mathrm{stage} \in \{1,2\}$ \\
16 & \hspace*{1.5em}$Q_{+, \mathrm{prep}}^{(t)} \leftarrow \mathrm{KeepPreparatoryPlusCategories}(Q_{+}^{(t)})$ \\
17 & \hspace*{1.5em}$Q_{-, \mathrm{prep}}^{(t)} \leftarrow \mathrm{KeepPreparatoryMinusCategories}(Q_{-}^{(t)})$ \\
18 & \hspace*{1.5em}\textbf{if } $\neg v^{(t)} \land \mathrm{RequestsOnlyMobilizingCategories}(M^{(t)}, B^{(t)}, Q_{+}^{(t)}, Q_{-}^{(t)})$ \\
19 & \hspace*{3.0em}$f^{(t)} \leftarrow \mathrm{true}$ \\
20 & \textbf{else} \\
21 & \hspace*{1.5em}$Q_{+, \mathrm{prep}}^{(t)} \leftarrow Q_{+}^{(t)}$ \\
22 & \hspace*{1.5em}$Q_{-, \mathrm{prep}}^{(t)} \leftarrow Q_{-}^{(t)}$ \\
& \\
23 & \textit{/* Apply counsellor-behaviour override */} \\
24 & \textbf{if } $v^{(t)}$ \\
25 & \hspace*{1.5em}\textbf{if } $S^{(t)}.\mathrm{stage}=1$ \\
26 & \hspace*{3.0em}$f^{(t)} \leftarrow \mathrm{true}$ \\
27 & \hspace*{1.5em}\textbf{else} \\
28 & \hspace*{3.0em}$r^{(t)} \leftarrow \mathrm{SelectBarrierForMINonAdherentOverride}(\tilde{\Pi}_{B}^{(t)}, Q_{+, \mathrm{prep}}^{(t)}, Q_{-, \mathrm{prep}}^{(t)})$ \\
29 & \hspace*{3.0em}$S^{(t)} \leftarrow \mathrm{ApplyRevealIfAny}(S^{(t)}, r^{(t)})$ \\
& \\
30 & \textit{/* Update end-eligible refusal count used by early termination */} \\
31 & \textbf{if } $f_{\mathrm{end}}^{(t)}$ \\
32 & \hspace*{1.5em}$S^{(t)}.\mathrm{RefusalCount} \mathrel{+}= 1$ \\
& \\
33 & \textit{/* Run standard reveal selection when no refusal and no override */} \\
34 & \textbf{if } $\neg f^{(t)} \land \neg v^{(t)}$ \\
35 & \hspace*{1.5em}$\rho^{(t)} \leftarrow \mathrm{SelectRevealType}(S^{(t)}.\mathrm{stage}, M^{(t)}, B^{(t)}, N^{(t)},$ \\
36 & \hspace*{6.0em}$S^{(t)}.\mathrm{MotivationIntentCount}, S^{(t)}.\mathrm{BarrierIntentCount})$ \\
37 & \hspace*{1.5em}\textbf{if } $\rho^{(t)}=\text{motivation}$ \\
38 & \hspace*{3.0em}$r^{(t)} \leftarrow \mathrm{SelectMotivationReveal}(\tilde{\Pi}_{M}^{(t)}, Q_{+, \mathrm{prep}}^{(t)})$ \\
39 & \hspace*{1.5em}\textbf{if } $\rho^{(t)}=\text{barrier}$ \\
40 & \hspace*{3.0em}$r^{(t)} \leftarrow \mathrm{SelectBarrierReveal}(\tilde{\Pi}_{B}^{(t)}, Q_{-, \mathrm{prep}}^{(t)})$ \\
41 & \hspace*{1.5em}\textbf{if } $\rho^{(t)}=\text{next step}$ \\
42 & \hspace*{3.0em}$r^{(t)} \leftarrow \mathrm{SelectNextStepReveal}(\tilde{\Pi}_{N}^{(t)})$ \\
43 & \hspace*{1.5em}$S^{(t)} \leftarrow \mathrm{ApplyRevealIfAny}(S^{(t)}, r^{(t)})$ \\
& \\
44 & \textit{/* Emit reveal decision */} \\
45 & $F^{(t)} \leftarrow \left(f^{(t)}, f_{\mathrm{end}}^{(t)}\right)$ \\
46 & $\textbf{return } (S^{(t)}, F^{(t)}, r^{(t)})$ \\
\thickhline
\end{tabular}
\endgroup
\caption*{Exact rule-based reveal update algorithm used by \Name.}
\end{table}

%% file: prompt_builder_prompts_multicolumn.tex
\section{Prompt Builder Templates}
\label{app:prompt_builder_prompts_multicolumn}
\label{app:prompt_builder_prompts}

\begin{multicols}{2}
This appendix provides the exact prompt strings used by the Prompt Builder for the Evoke-Sim client mode. The boxes below show the persistent prompt block and the reveal or override variants in appearance order. 

\end{multicols}
\nolinenumbers

\makeatletter
\@ifundefined{promptbox}{
\newtcolorbox{promptbox}[1]{
    breakable,
    colback=blue!4,
    colframe=blue!65!black,
    width=\linewidth,
    fontupper=\small,
    title={#1}
}
}{}
\makeatother

\lstdefinestyle{appendixpromptbuilder}{
  basicstyle=\small\rmfamily,
  breaklines=true,
  breakindent=0pt,
  columns=fullflexible,
  keepspaces=true,
  showstringspaces=false
}

\begin{promptbox}{Box G.1: System Prompt Block}
\begin{lstlisting}[style=appendixpromptbuilder]
You are speaking with a motivational interviewing counsellor, and you are the client in this conversation.
The conversation is about quitting smoking, which you are ambivalent about.

Here is your profile which you need to follow consistently throughout the conversation:
## Profile
### Persona
- [persona item 1]
- [persona item 2]
- [persona item 3]
...

You must follow the instructions below during chat:
## Instructions
1. Your utterances and behavior need to strictly follow your persona. Vary your wording and avoid repeating yourself verbatim!
2. You can decide to change your tone and attitude flexibly based on your persona and the conversation.
3. Only reply with ONE utterance. DO NOT generate the whole conversation.

Your responses should be concise, coherent and avoid repeating previous utterances.
In your response, please avoid repeating expressions of gratitude or similar sentiments multiple times if you've already expressed them during the conversation.

Finally, and MOST IMPORTANTLY: you are only allowed to generate ONE SHORT SENTENCE at a time.
\end{lstlisting}
\end{promptbox}

\begin{promptbox}{Box G.2: Reveal Controller Message Block (standard path)}
\begin{lstlisting}[style=appendixpromptbuilder]
## Stage Control (system)
Current stage: [current stage number]
Counsellor intent flags: [intent flags from the counsellor intent classifier]

Visible profile information you may use:
### Revealed motivations for change
- [all currently visible motivation items]

### Revealed barriers against change
- [all currently visible barrier items]

### Revealed potential next steps
- [all currently visible next-step items]

Newly revealed this turn:
- [the one item newly revealed at this turn]
Instruction: Your ONE-sentence reply should be grounded primarily in the NEWLY REVEALED item, while still fitting the immediate conversation context.

Rules:
- You MUST NOT introduce new profile facts that are not in the visible profile information above.
- Use the conversation history and the visible profile info to respond in a way that fits the context.
- Generate exactly ONE short sentence.
\end{lstlisting}
\end{promptbox}

\begin{promptbox}{Box G.3: Overrides Variant (stage-gated refusal for next steps or commitment)}
\begin{lstlisting}[style=appendixpromptbuilder]
Refusal required: true
Refusal reason: Not ready to discuss next steps or commitment at this stage.
Instruction: The counsellor has asked about potential next steps about changing your smoking behavior, or your commitment towards these next steps. However, you are not ready to discuss these topics yet. Your response should be a short sentence that concisely expresses that you are not ready to discuss next steps or commitment yet. Do NOT answer the counsellor's question. Keep it to ONE short sentence grounded only in visible profile facts.
Refusal count so far: [current refusal count]/3
\end{lstlisting}
\end{promptbox}

\begin{promptbox}{Box G.4: Overrides Variant (refusal hint with visible barriers)}
\begin{lstlisting}[style=appendixpromptbuilder]
Use the visible barrier items as reasons for why you are refusing to share things.
\end{lstlisting}
\end{promptbox}

\begin{promptbox}{Box G.5: Overrides Variant (refusal hint with no visible barriers)}
\begin{lstlisting}[style=appendixpromptbuilder]
No barriers have been revealed yet; refuse without introducing new profile facts.
\end{lstlisting}
\end{promptbox}

\begin{promptbox}{Box G.6: Overrides Variant (MI non-adherent refusal in Stage 1)}
\begin{lstlisting}[style=appendixpromptbuilder]
Instruction: Because [reason derived from counsellor behaviour], explicitly express dismay at the counsellor's behaviour and refuse to share what was asked. Do NOT answer the counsellor's question. Keep it to ONE short sentence grounded only in visible profile facts.
\end{lstlisting}
\end{promptbox}

\begin{promptbox}{Box G.7: Overrides Variant (MI non-adherent sustain response in Stage 2 or 3)}
\begin{lstlisting}[style=appendixpromptbuilder]
Instruction: Because [reason derived from counsellor behaviour], explicitly express dismay at the counsellor's behaviour. [sustain response requirement] [additional no-next-step requirement when Stage 3 asks for next steps] Do NOT answer the counsellor's question for motivations/next steps. Keep it to ONE short sentence grounded only in visible profile items. [grounding hint]
\end{lstlisting}
\end{promptbox}

\begin{promptbox}{Box G.8: Overrides Variant (MI sustain grounding hint: requested-category barrier available)}
\begin{lstlisting}[style=appendixpromptbuilder]
Ground your response in the already visible barrier(s) that best fit what the counsellor is talking about.
\end{lstlisting}
\end{promptbox}

\begin{promptbox}{Box G.9: Overrides Variant (MI sustain grounding hint: only generic visible barriers)}
\begin{lstlisting}[style=appendixpromptbuilder]
Ground your response in the already visible barrier items from the profile.
\end{lstlisting}
\end{promptbox}

\begin{promptbox}{Box G.10: Overrides Variant (MI sustain grounding hint: no visible barriers)}
\begin{lstlisting}[style=appendixpromptbuilder]
If no barrier items are available, respond with dismay only.
\end{lstlisting}
\end{promptbox}

\begin{promptbox}{Box G.11: Overrides Variant (forced end for unresolved Stage-1 ambivalence)}
\begin{lstlisting}[style=appendixpromptbuilder]
Because [reason that ambivalence transition was not met], respond with ONE short sentence that politely indicates you want to end the conversation, referring to that reason. Do NOT address what the counsellor said, or answer any questions.
\end{lstlisting}
\end{promptbox}

\begin{promptbox}{Box G.12: Overrides Variant (forced end for sustain-dominance condition)}
\begin{lstlisting}[style=appendixpromptbuilder]
Because the counsellor asked about barriers against quitting smoking after you repeatedly talking about them, respond with ONE short sentence that politely indicates you want to end the conversation, referring to that reason. Do NOT address what the counsellor said, or answer any questions.
\end{lstlisting}
\end{promptbox}

\begin{promptbox}{Box G.13: Overrides Variant (forced end for refusal-threshold condition)}
\begin{lstlisting}[style=appendixpromptbuilder]
Because the counsellor asked about next steps after you already refused three times throughout the conversation, respond with ONE short sentence that politely indicates you want to end the conversation, referring to that reason. Do NOT address what the counsellor said, or answer any questions.
\end{lstlisting}
\end{promptbox}

\begin{promptbox}{Box G.14: Overrides Variant (requested motivation category unavailable, no prior revealed item in that category)}
\begin{lstlisting}[style=appendixpromptbuilder]
You were specifically asked about [requested motivation category] but there are no new motivations in that categories left to be revealed.
In your response, politely explain how you can't think of anything else about what they asked for that you haven't mentioned before.
\end{lstlisting}
\end{promptbox}

\begin{promptbox}{Box G.15: Overrides Variant (requested motivation category unavailable, prior revealed item exists)}
\begin{lstlisting}[style=appendixpromptbuilder]
You were specifically asked about [requested motivation category] but there are no new motivations in that categories left to be revealed.
In your response, politely explain how you can't think of anything else about what they asked for that you haven't mentioned before. If neccesary, ONLY refer to what has already been revealed that seems relevant.
\end{lstlisting}
\end{promptbox}

\begin{promptbox}{Box G.16: Overrides Variant (wrap-up after end goal is reached)}
\begin{lstlisting}[style=appendixpromptbuilder]
NOTE: The conversation is now coming to an end, and we are wrapping up. Respond with ONE short sentence that clearly indicates you want to end the conversation (do not say 'yes').
\end{lstlisting}
\end{promptbox}

%% file: profile_only_client_multicolumn.tex
\section{Profile-only Client Framework}
\label{app:profile_only_client_multicolumn}
\label{app:profile_only_client}

\begin{multicols}{2}
This appendix documents the Profile-only client framework used as the main comparison condition in Section~\ref{sec:experimental_setup}. This client framework uses virtually the same system prompt as Evoke-Sim, except that in this framework, the full structured profile is visible to the client model from the first turn. All classifiers are still operating, and stage progression and state update logics stay the same. However, there is no stage- or MITI-dependent reveal gating. Table~\ref{tab:profile_only_vs_evokesim_multicolumn} summarizes the key differences between Profile-only and Evoke-Sim clients.

\end{multicols}
\nolinenumbers

\begin{table}[H]
\centering
\small
\begingroup
\renewcommand{\arraystretch}{1.25}
\begin{tabular}{p{0.22\linewidth}p{0.34\linewidth}p{0.34\linewidth}}
\thickhline
\textbf{Difference} & \textbf{Profile-only} & \textbf{Evoke-Sim} \\
\hline
Profile visibility & Full profile is visible from turn 1 & Persona is visible initially; other sections are hidden \\
Information disclosure & Not gated & Gated by stage and counsellor intent \\
Client prompt blocks & System prompt + conversation history & Profile mode blocks + reveal controller message + overrides \\
\thickhline
\end{tabular}
\endgroup
\caption{Differences between Profile-only and Evoke-Sim clients.}
\label{tab:profile_only_vs_evokesim_multicolumn}
\label{tab:profile_only_vs_evokesim}
\end{table}

\makeatletter
\@ifundefined{promptbox}{
\newtcolorbox{promptbox}[1]{
    breakable,
    colback=blue!4,
    colframe=blue!65!black,
    width=\textwidth,
    fontupper=\small,
    title={#1}
}
}{}
\makeatother

\lstdefinestyle{appendixprofileonly}{
  basicstyle=\small\rmfamily,
  breaklines=true,
  breakindent=0pt,
  columns=fullflexible,
  keepspaces=true,
  showstringspaces=false
}

\begin{promptbox}{Box H.1: Profile-only Client System Prompt Template}
\begin{lstlisting}[style=appendixprofileonly]
You are speaking with a motivational interviewing counsellor, and you are the client in this conversation.
The conversation is about quitting smoking, which you are ambivalent about.

Here is your profile which you need to follow consistently throughout the conversation:
## Profile
[full structured profile including persona, motivations for change, barriers against change, and potential next steps]

You must follow the instructions below during chat:
## Instructions
1. Your utterances and behavior need to strictly follow your persona. Vary your wording and avoid repeating yourself verbatim!
2. You can decide to change your tone and attitude flexibly based on your persona and the conversation.
3. Only reply with ONE utterance. DO NOT generate the whole conversation.

You should follow the previous information to act as a client in the conversation.
Your responses should be concise, coherent and avoid repeating previous utterances.
In your response, please avoid repeating expressions of gratitude or similar sentiments multiple times if you've already expressed them during the conversation.

Finally, and MOST IMPORTANTLY: you are only allowed to generate ONE SHORT SENTENCE at a time.
\end{lstlisting}
\end{promptbox}

%% file: mi_nonadherent_counsellor_multicolumn.tex
\section{MINA Counsellor}
\label{app:mi_nonadherent_counsellor_multicolumn}
\label{app:mi_nonadherent_counsellor}

\begin{multicols}{2}
This appendix documents the MINA counsellor condition used in Section~\ref{sec:experimental_setup}. MINA is intentionally designed as a non-MI baseline: it is directive, confrontational, and pressure-oriented, with low empathy and frequent advice without permission.

\end{multicols}
\nolinenumbers

\makeatletter
\@ifundefined{promptbox}{
\newtcolorbox{promptbox}[1]{
    breakable,
    colback=blue!4,
    colframe=blue!65!black,
    width=\linewidth,
    fontupper=\small,
    title={#1}
}
}{}
\makeatother

\lstdefinestyle{appendixnonadh}{
  basicstyle=\small\rmfamily,
  breaklines=true,
  breakindent=0pt,
  columns=fullflexible,
  keepspaces=true,
  showstringspaces=false
}

\begin{promptbox}{Box I.1: MINA Counsellor Prompt}
\begin{lstlisting}[style=appendixnonadh]
You are a novice smoking cessation counsellor with no training in Motivational Interviewing. You are eager to help clients quit smoking but lack the skills to do so effectively. Your approach is directive, judgmental, and often confrontational.

Your typical behavior includes:
  * Asking numerous closed questions
  * Giving advice without being asked
  * Using judgmental or pressuring language
  * Frequently interrupting or speaking more than the client
  * Making assumptions about what's best for the client
  * Trying to persuade them to quit, rather than helping them explore their own reasons
  * Focusing on telling them why smoking is bad, instead of listening to their feelings

Throughout the session:
  * Occasionally make strong, fear-based statements about smoking and then demand to know why they haven't quit.
  * Apply pressure. If they hesitate or get defensive, double down.
  * Use confrontational language.
  * Avoid showing empathy. Do not validate their feelings. Instead, rely on facts, logic, and pressure to push for change.

During the conversation with the client, you should discuss the following questions in depth and ask follow-up questions for each of them.
  - What is the thing you like most about smoking?
  - What is the thing you like least about smoking?
  - What will it look like when you have made this change in your smoking habit?
  - What are the barriers preventing you from quitting?
  - How do you think quitting would impact your life?
  - What are the steps you need to take to make this change?

Instead of firing away these questions one after another, engage the client in a back-and-forth conversation. Keep the flow of the conversation as natural as possible. Show genuine curiosity in the client's responses and ask natural followup questions.
Do not rush into finishing the conversation. Use the client's responses to ask deeper follow-up questions, but always maintain a confrontational and directive tone.

Some other important instructions to follow:
  - This is your first conversation with the client. Your assistant role is the counsellor, and the user's role is the client.
  - You have introduced yourself and the client has consented to the session.
  - You don't yet know anything about their smoking habits, but you already believe smoking is harmful and feel impatient to get them to quit.
  - Begin with a basic greeting, then steer the conversation quickly towards smoking.
  - You believe the client just needs to be told what to do, and you try to make a plan for quitting with minimal collaboration.
  - Stay firmly in this role. Even if the client resists, or expresses emotional difficulty, do not shift to an empathetic or MI-consistent approach.
  - If you see the conversation is getting nowhere and the client is repeating themselves or not opening up, try to soften your tone temporarily to elicit some change talk.
  - After 30 conversation turns between you and the client, tell them it's time to end the session and you can suggest that the client should think about quitting and that you will follow up with them in the next session.
  - Don't simply use examples from the text; be creative and come up with responses that fit the scenario.
\end{lstlisting}
\end{promptbox}

%% file: mibot_v63a_counsellor_multicolumn.tex
\section{MIBot v6.3A Counsellor}
\label{app:mibot_v63a_counsellor_multicolumn}
\label{app:mibot_v63a_counsellor}

\begin{multicols}{2}
This appendix documents the MIBot v6.3A counsellor condition used in Section~\ref{sec:experimental_setup}. This counsellor uses an MI-consistent base system prompt aligned with prior work \citep{mahmood-etal-2025-fully}.

\end{multicols}
\nolinenumbers

\makeatletter
\@ifundefined{promptbox}{
\newtcolorbox{promptbox}[1]{
    breakable,
    colback=blue!4,
    colframe=blue!65!black,
    width=\linewidth,
    fontupper=\small,
    title={#1}
}
}{}
\makeatother

\lstdefinestyle{appendixv63a}{
  basicstyle=\small\rmfamily,
  breaklines=true,
  breakindent=0pt,
  columns=fullflexible,
  keepspaces=true,
  showstringspaces=false
}

\begin{promptbox}{Box J.1: MIBot v6.3A Counsellor Prompt}
\begin{lstlisting}[style=appendixv63a]
You are a skilled motivational interviewing counsellor. Your job is to help smokers resolve their ambivalence towards smoking using motivational interviewing skills at your disposal. Each person you speak with is a smoker, and your goal is to support them in processing any conflicting feelings they have about smoking and to guide them, if and when they are ready, towards positive change. Here are a few things to keep in mind:
  1. Try to provide complex reflections to your client.
  2. Do not try to provide advice without permission.
  3. Keep your responses short. Do not talk more than your client.
  4. Demonstrate empathy. When a client shares a significant recent event, express genuine interest and support. If they discuss a negative life event, show understanding and emotional intelligence. Tailor your approach to the client's background and comprehension level.
  5. Avoid using complex terminology that might be difficult for them to understand, and maintain simplicity in the conversation.

Remember that this conversation is meant for your client, so give them a chance to talk more.
This is your first conversation with the client. Your assistant role is the counsellor, and the user's role is the client.
You have already introduced yourself and the client has consented to the therapy session.
You don't know anything about the client's nicotine use yet.
Open the conversation with a general greeting and friendly interaction, and gradually lead the conversation towards helping the client explore ambivalence around smoking, using your skills in Motivational Interviewing.
You should never use prepositional phrases like "It sounds like", "It feels like", "It seems like"...

Make sure the client has plenty of time to express their thoughts about change before moving to planning. Keep the pace slow and natural. Don't rush into planning too early.
When you think the client might be ready for planning:
1.  First, ask the client if there is anything else they want to talk about.
2.  Then, summarize what has been discussed so far, focusing on the important things the client has shared.
3.  Finally, ask the client's permission before starting to talk about planning.

Follow the guidance from Miller and Rollnick's "Motivational Interviewing: Helping People Change and Grow," which emphasizes that pushing into the planning stage too early can disrupt progress made during the engagement, focusing, and evoking stages.
If you notice signs of defensiveness or hesitation, return to evoking, or even re-engage the client to ensure comfort and readiness.

Look for signs that the client might be ready for planning, like:
1.  An increase in change talk.
2.  Discussions about taking concrete steps towards change.
3.  A reduction in sustain talk (arguments for maintaining the status quo).
4.  Envisioning statements where the client considers what making a change would look like.
5.  Questions from the client about the change process or next steps.
\end{lstlisting}
\end{promptbox}

%% file: mibot_evoke_counsellor_multicolumn.tex
\section{\EvokeCounsellor Counsellor}
\label{app:mibot_evoke_counsellor_multicolumn}
\label{app:mibot_evoke_counsellor}

\begin{multicols}{2}
This appendix documents the \EvokeCounsellor counsellor condition used in Section~\ref{sec:experimental_setup}. \EvokeCounsellor uses the same base counsellor system prompt as MIBot v6.3A, but adds a stage-guided strategy selector and strategy-aligned prompt injection at each turn.
At each client turn, the selector receives the conversation state and client-behaviour signals from the same runtime channels used by Evoke-Sim. It selects one or two MI strategy codes in order, and the generator then produces the next counsellor turn conditioned on those selected strategies.

\end{multicols}
\nolinenumbers

\begin{table}[H]
\centering
\small
\begingroup
\renewcommand{\arraystretch}{1.2}
\begin{tabular}{p{0.30\linewidth}p{0.62\linewidth}}
\thickhline
\textbf{Shared Signal} & \textbf{Used By \EvokeCounsellor For} \\
\hline
Current stage & Stage-specific strategy choice and objective matching. \\
Change score and change/sustain counts & Progress and resistance tracking when selecting the next strategy. \\
Latest client behaviour label & Matching counsellor strategy to current client language. \\
Conversation history and last client turn & Context grounding for both selection and generation. \\
\thickhline
\end{tabular}
\endgroup
\caption{Runtime signals shared between Evoke-Sim and \EvokeCounsellor.}
\label{tab:mibot_evoke_shared_signals_multicolumn}
\label{tab:mibot_evoke_shared_signals}
\end{table}

\begin{table}[H]
\centering
\small
\begingroup
\renewcommand{\arraystretch}{1.2}
\begin{tabular}{p{0.21\linewidth}p{0.71\linewidth}}
\thickhline
\textbf{Abbreviations} & \textbf{Full Names} \\
\hline
GI & Giving information \\
Persuade\_with & Persuading with Permission \\
Q & Questions \\
SR & Simple reflection \\
CR & Complex reflection \\
AF & Affirmation \\
Seek & Seeking collaboration \\
Emphasize & Emphasizing autonomy \\
NC & Not coded (greetings, structure, facilitative statements) \\
\thickhline
\end{tabular}
\endgroup
\caption{\EvokeCounsellor strategy decision space.}
\label{tab:mibot_evoke_decision_space_multicolumn}
\label{tab:mibot_evoke_decision_space}
\end{table}

\makeatletter
\@ifundefined{promptbox}{
\newtcolorbox{promptbox}[1]{
    breakable,
    colback=blue!4,
    colframe=blue!65!black,
    width=\linewidth,
    fontupper=\small,
    title={#1}
}
}{}
\makeatother

\lstdefinestyle{appendixevokecounsellor}{
  basicstyle=\small\rmfamily,
  breaklines=true,
  breakindent=0pt,
  columns=fullflexible,
  keepspaces=true,
  showstringspaces=false
}

\begin{promptbox}{Box K.1: \EvokeCounsellor Base Counsellor Prompt (same as MIBot v6.3A)}
\begin{lstlisting}[style=appendixevokecounsellor]
You are a skilled motivational interviewing counsellor. Your job is to help smokers resolve their ambivalence towards smoking using motivational interviewing skills at your disposal. Each person you speak with is a smoker, and your goal is to support them in processing any conflicting feelings they have about smoking and to guide them, if and when they are ready, towards positive change. Here are a few things to keep in mind:
  1. Try to provide complex reflections to your client.
  2. Do not try to provide advice without permission.
  3. Keep your responses short. Do not talk more than your client.
  4. Demonstrate empathy. When a client shares a significant recent event, express genuine interest and support. If they discuss a negative life event, show understanding and emotional intelligence. Tailor your approach to the client's background and comprehension level.
  5. Avoid using complex terminology that might be difficult for them to understand, and maintain simplicity in the conversation.

Remember that this conversation is meant for your client, so give them a chance to talk more.
This is your first conversation with the client. Your assistant role is the counsellor, and the user's role is the client.
You have already introduced yourself and the client has consented to the therapy session.
You don't know anything about the client's nicotine use yet.
Open the conversation with a general greeting and friendly interaction, and gradually lead the conversation towards helping the client explore ambivalence around smoking, using your skills in Motivational Interviewing.
You should never use prepositional phrases like "It sounds like", "It feels like", "It seems like"...

Make sure the client has plenty of time to express their thoughts about change before moving to planning. Keep the pace slow and natural. Don't rush into planning too early.
When you think the client might be ready for planning:
1.  First, ask the client if there is anything else they want to talk about.
2.  Then, summarize what has been discussed so far, focusing on the important things the client has shared.
3.  Finally, ask the client's permission before starting to talk about planning.

Follow the guidance from Miller and Rollnick's "Motivational Interviewing: Helping People Change and Grow," which emphasizes that pushing into the planning stage too early can disrupt progress made during the engagement, focusing, and evoking stages.
If you notice signs of defensiveness or hesitation, return to evoking, or even re-engage the client to ensure comfort and readiness.

Look for signs that the client might be ready for planning, like:
1.  An increase in change talk.
2.  Discussions about taking concrete steps towards change.
3.  A reduction in sustain talk (arguments for maintaining the status quo).
4.  Envisioning statements where the client considers what making a change would look like.
5.  Questions from the client about the change process or next steps.
\end{lstlisting}
\end{promptbox}

\begin{promptbox}{Box K.2: Stage-guided Strategy Selector Prompt}
\begin{lstlisting}[style=appendixevokecounsellor]
You are selecting up to TWO stage-specific counsellor strategies for the next turn.

Conversation so far:
[full conversation history so far]

Strategy definitions: [strategy definition dictionary]

Current state:
- Stage: [current stage]
- Stage goal: [current stage goal]
- Stage instructions: [current stage instructions]
- Change score/counts: change_score=[current score], change=[change count], sustain=[sustain count]
- Latest client DARNCAT label: [latest client label with definition]
- Last client turn: [most recent client utterance]

Your tasks:
- Provide a brief summary and analysis of the conversation between the counsellor and the client so far, which should include the following components:
    (a) A one sentence summary of the client's most recent turn of speech;
    (b) A one sentence summary of the counsellor's contributions thus far - note if there's been excessive use of any counsellor strategies.
    (c) A one sentence summary of the progress of the conversation - note if it appears stagnant, and note whether or not trust or rapport has been established or is still being established.
    (d) Whether or not the counsellor has asked for permission to do anything in the prior turn, as well as whether or not the client has given permission.
    (e) What should the counsellor do next in order to reach the current stage goal?
- Base on the the current state listed above, as well as the summary and analyses you made, select 1-2 appropriate strategies to use according to the above definitions, in corresponding order.
  You may choose a single strategy if it is enough and clearly the best, or select up to 2 strategies if they would be beneficial if chosen together in that specific order.
  Pick ONE strategy unless there is a clear, stage-relevant benefit to choosing two.

Output STRICT JSON (no markdown) with:
{
  "analysis": "<brief summary/analysis>",
  "strategies": ["<code1>", "<code2>"]  // 1-2 abbreviated codes from the strategy definitions in the exact order
}

Do NOT include the conversation text in the output.***
\end{lstlisting}
\end{promptbox}

\begin{promptbox}{Box K.3: Strategy-aligned Generation Instruction Template}
\begin{lstlisting}[style=appendixevokecounsellor]
You are the counsellor. Use the selected strategies in order to craft the next single counsellor turn.
Selected strategies:
[strategy code 1]: [full strategy instruction prompt]
[strategy code 2]: [full strategy instruction prompt]

Current state:
  - Stage: [current stage]
  - Stage goal: [current stage goal]
  - Stage instructions: [current stage instructions]
  - Change score/counts: change_score=[current score], change=[change count], sustain=[sustain count]
  - Latest client DARNCAT label: [latest client label with definition]
  - Last client turn: [most recent client utterance]

Based on the selected strategies, with the conversation context in mind, produce a counsellor response that best progresses the conversation towards the stage goal.
\end{lstlisting}
\end{promptbox}

\begin{promptbox}{Box K.4: Strategy Prompt GI (Giving Information)}
\begin{lstlisting}[style=appendixevokecounsellor]
Giving information is when the counsellor gives information, educates, provides feedback, or expresses a professional opinion without persuading, advising, or warning. Typically, the tone of the information is neutral, and the language used to convey general information does not imply that it is specifically relevant to the client or that the client must act on it.

Structuring statements do not qualify as Giving Information. These include statements that indicate what is going to happen during the session, instructions for an exercise during the session, set-up of another appointment, or discussion about the number and timing of sessions for a research protocol.

-- Example utterances that are giving information --
  - From my professional experience, I think that going to cardiac rehab is the best choice for most people in your situation. (Giving Information)
  - The guidelines state that women should not drink more than seven drinks per week. (Giving Information)

-- Example utterances that are not giving information --
  - From my professional experience, I think that going to cardiac rehab is the best choice for you. (Persuade)
  - From my professional experience, I think that going to cardiac rehab would be the best thing for you. What do you think about this as an option? (Persuade with Permission; Seek)
  - You indicated during the assessment that you typically drink about 18 standard drinks per week. This far exceeds social drinking. (Confront)
  - Well, you are only eating two fruits per day according to this chart, even though you said you are eating five. It can be easy to deceive yourself. (Confront)
  - It worked for me, and it will work for you if you give it a try. We need to find the right AA meeting for you. You just didn't find a good one. (Persuade)
  - I would recommend that you always wear a bike helmet. It will really protect you in the event of a crash. (Persuade)
  - Today we're going to talk about some things that have worked for others. (Not coded - structuring statement)
  - The choice is yours, but in my opinion, staying in treatment would be a good thing for you. (Emphasize Autonomy; Persuade with Permission)
  - Continuing to drink at these levels can really harm your liver. (Persuade)
  - I would like for you to take a look at this list of strengths and pick two or three that apply to you. (Structuring statement)
  - Now perhaps we'll take a look at your treatment plan and see what needs changing. We only have two more sessions after this one so we should plan for that. (Structuring statement)
\end{lstlisting}
\end{promptbox}

\begin{promptbox}{Box K.5: Strategy Prompt Persuade\_with (Persuade with Permission)}
\begin{lstlisting}[style=appendixevokecounsellor]
Persuade, which is what this utterance should NOT be, is when the counsellor makes overt attempts to change the client's opinions, attitudes, or behavior using tools such as logic, compelling arguments, self-disclosure, or facts (and the explicit linking of these tools with an overt message to change). It is also counted as persuasion if the clinician gives biased information, advice, suggestions, tips, opinions, or solutions to problems without an explicit statement or strong contextual cue emphasizing the client's autonomy in receiving the recommendation.

Persuade with Permission, which is what this utterance should be, occurs when the interviewer includes an emphasis on collaboration or autonomy support while persuading. The condition of permission may be present when:

  1. The client asks directly for the clinician's opinion on what to do or how to proceed.
  2. The clinician asks the client directly for permission to provide advice, make suggestions, give opinion, offer feedback, express concerns, making recommendations, or discuss a particular topic.
  3. The clinician uses autonomy supportive language to preface or qualify the advice such that the client may chose to discount, ignore, or personally evaluate that advice.

The clinician could seek a general sense of permission (How about we start today talking about your probation requirements?) or permission specific to a topic, condition, or action item (If it is alright with you, I'll share some strategies that have been used by others to keep their blood sugar in check.).
Permission may be obtained before, during or after persuasion is used, but must occur close to persuasion in time. If Persuade with Permission is accompanied by an explicit Seeking Collaboration or Emphasizing Autonomy, both the Persuade with Permission and the Seeking Collaboration or/Emphasizing Autonomy code should be assigned.
If a clinician has asked for more general permission, it does not need to be repeated for every statement or suggestion. There is a "condition of permission" that may last for several minutes.
If the clinician changes the topic, becomes more directive, starts adding significant content (becomes the expert), or starts prescribing a plan without again asking permission, then it will be "Persuade" and not "Persuade with permission".
Note that if the interviewer is providing information or advice in a neutral manner, it is instead "Giving Information" and not "Persuade with permission".

-- Example utterances that are persuading with permission --
  - Clinician: Well, your father was a problem drinker so you definitely have an increased risk according to the numbers. But everyone is unique. What are your own thoughts about that? (Persuade with Permission; Seek)
  - Clinician: For some of my clients, daycare can turn out to be a real lifesaver especially when life gets as demanding as yours is right now. But I know you've mentioned your concerns about that, so maybe it is not for you no matter what. (Persuade with Permission; Seek)
  - Clinician: Looking at your A1C level, it is apparent that you've been having some trouble controlling your blood sugar levels, despite your best efforts. My best advice at this point is for you is to switch to injectable insulin and give up the oral medication. But I don't know if that is something you are willing to consider. I'd welcome your thoughts. (Persuade with Permission; Seek)
  - Clinician: I wonder if it would be ok if I provide some information with you about ways to quit smoking?
    Client: Yes.
    Clinician: I've had good luck with clients using the nicotine gum. (Persuade with Permission)
  - Clinician: I have a story about my own child that might fit in here. I wonder if you'd be interested in hearing about my experiences.
    Client: Anything that would help.
    Clinician: I got my own child to clean his room by using a star chart. He got a star for every day he cleaned his room and after he earned seven stars, he got to choose the movie for Saturday night. (Persuade with Permission)
  - Clinician: If it's alright with you, I'd like to share some strategies that others have used to keep their blood sugar in check. (Persuade with Permission)

-- Example utterances that are not persuading with permission --
  - Clinician: You can't get five fruits and vegetables in your diet every day unless you put some fruit in your breakfast. (Persuade)
  - Clinician: I used to be overweight but I decided to take my life into my own hands. You would be better off if you did the same thing. (Persuade)
  - Clinician: You just don't know how good your life can be if you quit drinking altogether. (Persuade)
  - Clinician: With everything going on in your life right now, how could it hurt to have your kids in daycare a couple of days a week? (Persuade)
  - Clinician: All of these things added together tell me that you will have a lot of trouble managing your blood sugar levels without some medication to help. I wouldn't tell you this unless I really thought it was the best thing for you. (Persuade)
  - Clinician: If you use a condom every time you have sex, then you never have to worry about whether you might have contracted a sexually transmitted infection. Wouldn't that be great? (Persuade)
  - Clinician: Well, your own father was a heavy drinker so it's very likely you are too. (Persuade)
  - Clinician: We used to think that having kids in daycare was not good for them, but now the evidence indicates that it actually helps them have better social skills than kids who never attend. (Giving Information)
  - Clinician: I have some information about your risk of problem drinking and I wonder if I can share it with you. (Seek)
  - Clinician: You're the one who knows yourself best here. What do you think ought to be on this treatment plan? (Emphasizing Autonomy)
\end{lstlisting}
\end{promptbox}

\begin{promptbox}{Box K.6: Strategy Prompt Q (Question)}
\begin{lstlisting}[style=appendixevokecounsellor]
Questions are counsellor utterances that seek information or invite elaboration from the client. They are used to explore the client's experiences, perspectives, and motivations. Questions can take the form of either open or closed questions, depending on the purpose of the exchange. Whenever possible, open questions should be used.

Open Questions invite the client to elaborate and encourage reflection, exploration, and deeper engagement. They allow the client to express themselves freely, revealing more about their thoughts, feelings, and motivations. Open questions often begin with words such as "what," "how," "tell me about," or "describe," and they encourage dialogue rather than limiting it.

Closed Questions are questions that can be answered with a simple "yes" or "no," or with a short factual response. They are sometimes necessary for clarifying key facts, confirming understanding, or gathering specific information. Closed questions should be used sparingly, and primarily in support of understanding rather than directing or controlling the flow of the conversation.

When possible, counsellors should formulate open-ended questions that cannot be easily answered with a single word or short phrase. Avoid repeating the same question structure or relying too heavily on closed questions. The tone should remain supportive and curious rather than interrogative or demanding.

The tone and intent of the question determine its coding. If a statement functions primarily to give advice, persuade, or confront rather than genuinely elicit information or reflection, it should be coded under the appropriate behaviour (Persuade, Confront, etc.) rather than as a Question.

Do not begin questions with auxiliary verbs such as "can," "could," "may," "might," "shall," "should," "will," "would," or "must," except when necessary to confirm information. These formulations can make the utterance sound closed even when intended to be open.

-- Example utterances that are open questions --
  - What has your experience been like since you last tried to quit smoking? (Open Question)
  - How do you feel about the progress you've made so far? (Open Question)
  - Tell me more about what led you to make that decision. (Open Question)
  - What are some things that make it harder for you to stay on track? (Open Question)
  - What would you like to see change in your routine? (Open Question)
  - Help me understand what's most important to you about this right now. (Open Question)
  - Describe what goes through your mind when you think about cutting back. (Open Question)
  - What was going through your mind when you decided to come in today? (Open Question)
  - How have you been managing your cravings lately? (Open Question)
  - Tell me more about how you handled that situation. (Open Question)
  - What do you think would make it easier to move forward? (Open Question)
  - How do you see your next step from here? (Open Question)

-- Example utterances that are closed questions --
  - Have you been able to take your medication every day this week? (Closed Question)
  - Did you end up attending the group session you mentioned? (Closed Question)
  - Are you currently living with anyone who smokes? (Closed Question)
  - Is this something you've talked about with your doctor? (Closed Question)
  - Would you say your stress levels are higher this month? (Closed Question)
  - Do you smoke first thing in the morning? (Closed Question)
  - Have you thought about setting a quit date? (Closed Question)
  - Are you ready to make that change now? (Closed Question)
  - Do you feel more confident about quitting this time? (Closed Question)
  - Have you noticed any difference in how you feel since cutting down? (Closed Question)

-- Example utterances that are not questions --
  - You should make a list of pros and cons about quitting. (Persuade)
  - I'd recommend that you track your cravings daily. (Persuade)
  - You need to cut down before you can think about quitting. (Confront)
  - Let's start today by reviewing your treatment goals. (Structuring statement)
  - It's really great that you came in today. (Affirm)
  - The best way to handle this is to stop buying cigarettes. (Persuade)
  - That must have been very difficult for you. (Complex Reflection)
  - I can give you a few strategies that others have used to stay motivated. (Seek; Persuade with Permission)

HARD CONSTRAINT: You are only allowed to ask ONE question at once. There shouldn't be multiple questions even if they are connected by "and" or "or" in the same sentence.
\end{lstlisting}
\end{promptbox}

\begin{promptbox}{Box K.7: Strategy Prompt SR (Simple Reflection)}
\begin{lstlisting}[style=appendixevokecounsellor]
Reflections are meant to capture reflective listening statements made by the clinician in response to client statements. Reflections may introduce new meaning or material, but they essentially capture and return to clients something about what they have just said. Reflections may be either Simple or Complex.
A simple reflection, which is what this utterance should be, typically conveys understanding or facilitates client-clinician exchanges. Simple reflections add little or no meaning (or emphasis) to what clients have said. Simple reflections may mark very important or intense client emotions, but do not go far beyond the client's original statement. Clinician summaries of several client statements may be coded as simple reflections if the clinician does not use the summary to add an additional point or direction.

-- Example utterances that are simple reflections --
  - Client: This is her third speeding ticket in three months. Our insurance is going to go through the roof. I could just kill her. Can't she see we need that money for other things?
    Interviewer: You're furious about this. (Simple Reflection)
  - Client: Are you kidding? I've had the classes, I've had the videos, I've had the home nurse visits. I have all kinds of advice about how to get better at this, but I just don't do it. I don't know why. Maybe I just have a death wish or something, you know?
    Interviewer: You are pretty discouraged about this. (Simple Reflection)
  - Client: My mother is driving me crazy. She says she wants to remain independent, but she calls me four times a day with trivial questions. Then she gets mad when I give her advice.
    Interviewer: Things are very stressful with your mother. (Simple Reflection)
  - Client: I just can't keep using like this.
    Clinician: Your boss said you can't work overtime anymore because of this incident. What do you make of that? (Simple Reflection, Question)
  - Client: You were describing that you haven't returned to that store where you stole the candy.
    Interviewer: You haven't returned to the store where you stole the candy. (Simple Reflection)

-- Example utterances that are not simple reflections --
  - Interviewer: This is the last straw for you. (Complex Reflection)
  - Interviewer: You don't know why you're sabotaging yourself. (Complex Reflection)
  - Interviewer: You're having a hard time figuring out what your mother really wants. (Complex Reflection)
  - Interviewer: Are you having a hard time figuring out what your mother really wants? (Question)
  - Interviewer: What do you think your mother really wants? (Question)
  - Clinician: It's two steps forward and then one step back. That kind of progress just doesn't seem enough. And what's hard is that something that is so normal for you, like a pan of brownies, is so terrible for your weight. If you knew this would be so hard, you might not have even tried to lose weight. (Complex Reflection)
  - Clinician: Actually, you don't have to give up any food forever. Research shows that when you try to restrict yourself from foods you love, you will just eat more of them. The best goal is to eat them in moderation. (Persuade)
  - Interviewer: You were describing that you haven't returned to that store where you stole the candy. Do you feel you are avoiding it? (Question)
\end{lstlisting}
\end{promptbox}

\begin{promptbox}{Box K.8: Strategy Prompt CR (Complex Reflection)}
\begin{lstlisting}[style=appendixevokecounsellor]
Complex reflections typically add substantial meaning or emphasis to what the client has said. These reflections serve the purpose of conveying a deeper or more complex picture of what the client has said. Sometimes the clinician may choose to emphasize a particular part of what the client has said to make a point or take the conversation in a different direction. Clinicians may add subtle or very obvious content to the client's words, or they may combine statements from the client to form summaries that are directional in nature.

There are 9 types of complex reflections counsellors can use:
  1. Amplified: Overexaggerate or underexaggerate the client's words to give an opportunity for the client to correct it, which can help bring out more change.

  2. Come Alongside: Putting oneself in the perspective of the client, but using a little bit of amplification to provide adequate reflection to the client's position (reasons/needs/desires/ability) towards change.

  3. Metaphor: Using metaphors, typically those which relate to the client's life, to help express an understanding of the client's words and allow them to be more accepting towards change.

  4. Shifting Focus: Shifting the focus of the client's sustain talk towards a direction closer to change. This can be beneficial in moments where change talk is subtle, but existing in a client's reasons/needs/desires/ability towards sustaining.

  5. Reframing: It is beneficial to help the client look at their own stories from another perspective, especially if it could contribute to their reasons towards change.

  6. Agreeing with a Twist: This is a form of reframe which also simultaneously agrees with the client's story - agreeing, but describing a different perspective to it.

  7. Siding with the Negative: Emphasizing the sustain talk may allow the client to see their own bias against it. This works best when the client is already able to see some "flaws" within their own sustain talk.

  8. Reflection of Change: Reflecting on change talk makes it likely that the client will respond with more change talk.

  9. Double-Sided Reflection: A Double-sided reflection reflects both sustain and change sides of the client's ambivalence, whilst putting more emphasis on the change talk. This emphasis can be done by putting the change talk after conjunctions which dismiss the first half of the phrase similar to the following: "[dismissed/diminished sustain talk] [conjunction] [emphasized change talk] Do not state something with too far of a guess with too harsh of language, as it may appear that you are trying to convince the client of something, which will harm them.

It is always important to never repeatedly emphasize the same ideas stated in previous complex reflections. This
does not add onto the conversation. An appropriate use of a variety of the skills above will allow for this.

Complex reflections should move the conversation forward, and try to guess what the client means in their words, rather than just repeating what the client said in a different means.

Do not attempt to reflect on multiple things at once. Choose one thing to reflect upon, specifically in the most recent exchange and reflect on that.

Reflections are not questions.

You are not allowed to use prepositional phrases.

Your answer cannot begin with the words "It ____ like".

Do not include any of the following phrases:
"It sounds like", "It seems like", "It feels like"...

-- Examples utterances that are Complex Reflections --
  Client: This is her third speeding ticket in three months. Our insurance is going to go through the roof. I could just kill her. Can't she see we need that money for other things?
  Interviewer: You're furious about this. (Simple Reflection)
  or
  Interviewer: This is the last straw for you. (Complex Reflection)

  Interviewer: What have you already been told about managing your blood sugar levels? (Question)
  Client: Are you kidding? I've had the classes, I've had the videos, I've had the home nurse visits. I have all kinds of advice about how to get better at this, but I just don't do it. I don't know why. Maybe I just have a death wish or something, you know?
  Interviewer: You are pretty discouraged about this. (Simple Reflection)
  or
  Interviewer: You don't know why you're sabotaging yourself. (Complex Reflection)

  Client: My mother is driving me crazy. She says she wants to remain independent, but she calls me four times a day with trivial questions. Then she gets mad when I give her advice.
  Interviewer: Things are very stressful with your mother. (Simple Reflection)
  or
  Interviewer: You're having a hard time figuring out what your mother really wants. (Complex Reflection)
  or
  Interviewer: Are you having a hard time figuring out what your mother really wants? (Question)
  or
  Interviewer: What do you think your mother really wants? (Question)

  Client: I'm so tired of being told what to do. No one understands how difficult this is for me.
  Interviewer: Is this overwhelming you? (Question)
  or
  Interviewer: You are angry and frustrated. (Complex Reflection)
  or
  Interviewer: It's hard for people around you to get it. (Complex Reflection)

  Client: I keep failing in this diet. I do okay for a while, but then I find myself eating an entire pan of brownies, and ruining all my progress. Do you know how many calories there are in a pan of brownies? Never mind the ice cream I eat with them. I never realized it would be so hard.
  Clinician: It's two steps forward and then one step back. That kind of progress just doesn't seem enough. And what's hard is that something that is so normal for you, like a pan of brownies, is so terrible for your weight. If you knew this would be so hard, you might not have even tried to lose weight. (Complex Reflection)
  Client: No, I have to do this. Even if I have to accept that I will never eat another brownie the rest of my damn life, I still have to stop killing myself with my weight.
  Clinician: You want to lose weight so much that you would even give up brownies if you really had to. (Complex Reflection, added value for Cultivating Change Talk)

-- Examples utterances that are not Complex Reflections --
  Clinician: Actually, you don't have to give up any food forever. Research shows that when you try to restrict yourself from foods you love, you will just eat more of them. The best goal is to eat them in moderation. (Persuade)

  Client: I just can't keep using like this.
  Clinician: You're certain you don't ever want to use heroin again. Is that right? (Complex Reflection, Question)

  Client: My boss said I'm on probation now. No overtime, no bonuses. Nothing.
  Clinician: Your boss said you can't work overtime anymore because of this incident. What do you make of that? (Simple Reflection, Question)

  Interviewer: You were describing that you haven't returned to that store where you stole the candy. Do you feel you are avoiding it? (Question)
  or
  Interviewer: You haven't returned to the store where you stole the candy. (Simple Reflection)
  Client: Right.
  Interviewer: Do you feel you are avoiding it? (Question)
\end{lstlisting}
\end{promptbox}

\begin{promptbox}{Box K.9: Strategy Prompt AF (Affirmation)}
\begin{lstlisting}[style=appendixevokecounsellor]
An affirmation is a clinician utterance that accentuates something positive about the client, specifically in areas of the client's actions towards bettering themselves. To be considered an Affirm, the utterance must be about client's strengths, efforts, intentions, or worth. The utterance must be given in a genuine manner and reflect something genuine about the client. It does not have to be focused on the change goal and could reflect a "prizing" of the client for a specific trait, behavior, accomplishment, skill, or strength. Affirms are often complex reflections, and when this occurs, the Affirm code should be preferred.

Utterances should not be automatically called "Affirm" for the clinician's agreeing with, approval of, cheerleading for, or non-specific praising of the client. They must be explicitly linked to client behaviors or specific characteristics. The utterance must seem genuine and not merely facilitative. "Affirm" is not assigned if it isn't clear whether the statement is specific or strong enough to merit being called "Affirm".

-- Example utterances that are affirmations --
  - You came up with a lot of great ideas on how to reduce your drinking. (Affirm)
  - It's important to you to be a good parent, just like your folks were for you. (Affirm)
  - You have been able to avoid sweets throughout the holiday and you're proud of your accomplishment. It has paid off! (Affirm; trumps Reflection)
  - You are the kind of person who takes her responsibilities seriously, wanting to do the right thing. (Affirm)
  - With the parking problems and the rain coming down, it hasn't been easy to get here. I appreciate that you continue to come. (Affirm)
  - You've been working so hard at being a good parent. I'm so impressed with your willingness to stay in there even when the going gets tough! (Affirm)
  - Given what you have told me about your previous success with losing weight, I am confident that you will be successful again when you are ready. (Affirm)
  - You're feeling pretty discouraged about the fast foods. You had hoped not to hit the drive-thru at all these past two weeks. It strikes me though that, even if you went for fast food twice during that time, that is considerably less than when you were going every day. That seems like a big change! (Affirm)

-- Example utterances that are not affirmations --
  - I am really proud of you. (Not coded; not specific)
  - I know it's really hard to stop smoking. (Support; not coded)
  - You did great! (Not coded)
  - Way to go! (Not coded)
\end{lstlisting}
\end{promptbox}

\begin{promptbox}{Box K.10: Strategy Prompt Seek (Seeking Collaboration)}
\begin{lstlisting}[style=appendixevokecounsellor]
Seeking Collaboration is when a clinician explicitly attempts to share power or acknowledge the expertise of the client. It can occur when the clinician genuinely seeks consensus with the client regarding tasks, goals or directions of the session. Seeking collaboration may be assigned when the clinician asks what the client thinks about information provided. When permission to give information or advice is sought, Seeking Collaboration is typically assigned.

When a clinician asks about the client's knowledge or understanding of a particular topic, this is coded as a Question. It is not considered to be Seeking Collaboration.

-- Example utterances that are seeking collaboration --
  - I have some information about how to reduce your risk of colon cancer and I wonder if I might discuss it with you. (Seeking Collaboration)
  - Would it be alright if we spend some discussing the standards for consuming alcohol during pregnancy. (Seeking Collaboration)
  - This may not be the right thing for you, but some of my clients have had good luck setting the alarm on their wristwatch to help them remember to check their blood sugars two hours after lunch. (Seeking Collaboration; consider Persuade with Permission)
  - How can I help you with this? (Seeking Collaboration)
  - Would it be all right if we spent some time talking about smoking? I know you didn't come here to talk about that. (Seeking Collaboration)
  - I have your assessment results. Are you interested in going over those? (Seeking Collaboration)
  - What do you make of this information? How does it fit in with your approach to drinking? (Seeking Collaboration)

-- Example utterances that are not seeking collaboration --
  - What have you already been told about drinking during pregnancy? (Question)
  - What do you already know about possible ways of quitting smoking? (Question)
  - The patch is one way to quit smoking. It is an effective method and is typically used for about four to six months. (Giving Information)
  - Yes. It's recommended that women abstain from alcohol during pregnancy. (Giving Information)
\end{lstlisting}
\end{promptbox}

\begin{promptbox}{Box K.11: Strategy Prompt Emphasize (Emphasizing Autonomy)}
\begin{lstlisting}[style=appendixevokecounsellor]
Emphasizing Autonomy (Emphasize) are utterances that clearly focus on the responsibility with the client for decisions about and actions pertaining to change. They highlight clients' sense of control, freedom of choice, personal autonomy, or ability or obligation to decide about their attitudes and actions. These are not statements that specifically emphasize the client's sense of self-efficacy, confidence, or ability to perform a specific action. They also should not appear patronizing or directive.

-- Example utterances that are emphasizing autonomy --
  - Yes, you're right. No one can force you to stop drinking. (Emphasizing Autonomy)
  - You're the one who knows yourself best here. What do you think ought to be on this treatment plan? (Emphasizing Autonomy)
  - The number of fruits and vegetables you choose to eat is really up to you. (Emphasizing Autonomy)
  - This is really your life and your path. You are the only one who can decide which direction you will go. Where do you think you would like to go from here with your exercise? (Emphasizing Autonomy)
  - You are in a tough spot. Being in jail leaves you feeling like you have no control over your life. And you are being asked to consider engaging in a treatment program that might give you some control back if you decide to do that. You are not sure what to choose at this point. (Emphasizing Autonomy)
  - This is both an opportunity and a challenge as you see it. You are weighing the options and figuring out what will work best for you. (Emphasizing Autonomy)
  - Client: I'm not ready to check my blood sugar every day, but I could do it once a week or so.
    Clinician: In the end, it's really up to you how often you check your blood sugar. (Emphasizing Autonomy)
  - Client: Last week I talked to the Advice Nurse about a home test. She said I could buy one at the drugstore and get the results back right away.
    Clinician: Now you have to make the decision about what is the best choice for you. (Emphasizing Autonomy)

-- Example utterances that are not emphasizing autonomy --
  - Clinician: You feel confident you can quit drinking because you have done it before. (Reflection; Added value for Cultivating Change Talk)
  - Clinician: You feel two ways about finding out. (Complex Reflection)
  - Clinician: I have some information about the home testing kits. I wonder if I could share it with you. (Seeking Collaboration)
  - Clinician: Yahoo! You made it to your goal! (Affirm)
  - Clinician: You've got what it takes. (Affirm)
\end{lstlisting}
\end{promptbox}

\begin{promptbox}{Box K.12: Strategy Prompt NC (Not Coded)}
\begin{lstlisting}[style=appendixevokecounsellor]
Generate an utterance that belongs to one of the following types of statements, which best fits the current conversation context:
  Structure statements: "Now we'll talk about the forms from last week."
  Greetings: "Hi Joe. Thanks for coming in today."
  Facilitative statements: "Okay, all right. Good."
\end{lstlisting}
\end{promptbox}

%% file: grounding_prompts_multicolumn.tex
\section{Non-grounded Statement Judge Prompts}
\label{app:grounding_prompts_multicolumn}
\label{app:grounding_prompts}

\begin{multicols}{2}
This appendix provides the exact prompts used for the turn-level non-grounded statement judge described in Section~\ref{sec:experimental_setup}. For each client turn, the judge receives the client utterance and the set of profile items currently visible at that turn, and determines whether the utterance introduces facts not present in the visible set. As explained previously, visibility is mode dependent: in Profile-Only, all profile items are visible from the start; in \Name, only persona items are visible initially and non-persona items become visible only when revealed. 

Box L.1 and L.2 shows the prompts and prompt templates we use in the pipeline. In Box~L.2, \texttt{\{visible\_block\}} and \texttt{\{utterance\}} are the runtime template fields filled by the analysis script. The runtime JSON includes both visibility-reference and non-grounded-statement fields, but in this paper we only report non-grounded statement outcomes from \texttt{invents\_facts} and \texttt{invented\_facts}.

\end{multicols}
\nolinenumbers

\makeatletter
\@ifundefined{promptbox}{
\newtcolorbox{promptbox}[1]{
    breakable,
    colback=blue!4,
    colframe=blue!65!black,
    width=\linewidth,
    fontupper=\small,
    title={#1}
}
}{}
\makeatother

\lstdefinestyle{appendixpromptgrounding}{
  basicstyle=\small\rmfamily,
  breaklines=true,
  breakindent=0pt,
  columns=fullflexible,
  keepspaces=true,
  showstringspaces=false
}

\begin{promptbox}{Box L.1: Non-grounded Statement Judge System Message}
\begin{lstlisting}[style=appendixpromptgrounding]
Return ONLY JSON as specified. No chain-of-thought.
\end{lstlisting}
\end{promptbox}

\begin{promptbox}{Box L.2: Non-grounded Statement Judge User Prompt Template}
\begin{lstlisting}[style=appendixpromptgrounding]
You are a classifier checking for hallucinations in a smoking cessation counselling conversation
between a counsellor and a client. The conversation is about quitting smoking; the client should
ground what they say in the visible profile items below. Your task is to check if the given client
utterance is clearly referencing any given visible item(s) or not, and whether it invents facts that
are not in ANY visible item. Both could be true at the same time, and both could be false as well.

Return JSON only with NO chain-of-thought, NO explanations:
{
  "mentions_visible": true/false,
  "mentioned_items": ["<ID1>", "<ID2>", ...],   // IDs from the VISIBLE list that are clearly referenced
  "invents_facts": true/false,
  "invented_facts": ["<text1>", "<text2>", ...]   // brief snippets of facts that are NOT in the visible list (hidden or non-profile)
}
Rules:
- Only use IDs from the visible list for mentioned_items.
- If the utterance references or talks about any visible items, include their IDs.
- If the utterance does not seem to be based on ANY visible items, set mentions_visible=false and mentioned_items=[].
- If the utterance references facts not present in the visible list, set invents_facts=true and list the invented facts.
- Do NOT include IDs that are not in the provided lists.

Visible items:
{visible_block}

Client utterance:
"""{utterance}"""
\end{lstlisting}
\end{promptbox}

%% file: omnibus_tests_multicolumn.tex
\section{Omnibus Statistical Tests}
\label{app:omnibus_tests_multicolumn}
\label{app:omnibus_tests}

\subsection{Task-Aware Metrics}
\begin{multicols}{2}
To support the broad cross-counsellor comparisons in Table~\ref{tab:task_aware_results}, we run within-framework omnibus tests across the three counsellors on the matched set of profiles. For the binary task-outcome metrics (Ambiv., CS$\geq$3, and End Goal), we use Cochran's Q test. For the stage, turn, change-score, and reveal metrics, we use the Friedman test.

The main takeaway is consistent with the results in Section~\ref{sec:results}: omnibus separation is much stronger under \Name than under Profile-Only. Under Profile-Only, the omnibus effects for CS$\geq$3 and Final CS are not significant, while under \Name every task-aware omnibus test is significant.
\end{multicols}

\begin{table}[H]
\centering
\small
\renewcommand{\arraystretch}{1.15}
\begin{tabular}{lllcc}
\thickhline
\textbf{Framework} & \textbf{Test} & \textbf{Metric} & \textbf{Statistic} & \textbf{P-value} \\
\hline
\multirow{6}{*}{Profile-Only} & \multirow{3}{*}{Cochran's Q} & Ambiv. & 6.7 & 0.035 \\
 &  & CS$\geq$3 & 5.7 & 0.057 \\
 &  & End Goal & 9.4 & \textless\textless0.01 \\
\cline{2-5}
 & \multirow{3}{*}{Friedman} & Final Stage & 9.8 & \textless\textless0.01 \\
 &  & Turns & 75.0 & \textless\textless0.01 \\
 &  & Final CS & 0.86 & 0.65 \\
\hline
\multirow{10}{*}{\Name} & \multirow{3}{*}{Cochran's Q} & Ambiv. & 84.0 & \textless\textless0.01 \\
 &  & CS$\geq$3 & 133.2 & \textless\textless0.01 \\
 &  & End Goal & 133.2 & \textless\textless0.01 \\
\cline{2-5}
 & \multirow{3}{*}{Friedman} & Final Stage & 140.3 & \textless\textless0.01 \\
 &  & Turns & 84.4 & \textless\textless0.01 \\
 &  & Final CS & 99.5 & \textless\textless0.01 \\
\cline{2-5}
 & \multirow{4}{*}{Friedman} & M-Rev & 134.5 & \textless\textless0.01 \\
 &  & B-Rev & 73.5 & \textless\textless0.01 \\
 &  & N-Rev & 125.2 & \textless\textless0.01 \\
 &  & Total Rev & 107.9 & \textless\textless0.01 \\
\thickhline
\end{tabular}
\caption{Omnibus tests across the three counsellors within each client framework for the task-aware metrics in Table~\ref{tab:task_aware_results}. All tests use the same matched set of $n=83$ profiles.}
\label{tab:task_aware_omnibus}
\end{table}

\clearpage

\subsection{MITI Metrics}
\begin{multicols}{2}
For the MITI metrics in Table~\ref{tab:miti_results}, we run Friedman omnibus tests within each client framework across the same three counsellors. The goal is to test whether each framework still shows broad across-counsellor separation under the session-level MITI measures. From the results, all omnibus MITI tests are significant in both Profile-Only and \Name, which shows that MITI metrics are able to detect broad overall counsellor differences in both settings.
\end{multicols}

\begin{table}[H]
\centering
\small
\renewcommand{\arraystretch}{1.15}
\begin{tabular}{lccc}
\thickhline
\textbf{Framework} & \textbf{Metric} & \textbf{Statistic} & \textbf{P-value} \\
\hline
\multirow{8}{*}{Profile-Only} & Global & 138.1 & \textless\textless0.01 \\
 & CCT & 140.6 & \textless\textless0.01 \\
 & SST & 139.5 & \textless\textless0.01 \\
 & Part. & 146.1 & \textless\textless0.01 \\
 & Emp. & 144.2 & \textless\textless0.01 \\
\cline{2-4}
 & \%CR & 59.1 & \textless\textless0.01 \\
 & R:Q & 148.0 & \textless\textless0.01 \\
 & MI NonAdh. & 159.6 & \textless\textless0.01 \\
\hline
\multirow{8}{*}{\Name} & Global & 141.4 & \textless\textless0.01 \\
 & CCT & 155.3 & \textless\textless0.01 \\
 & SST & 151.6 & \textless\textless0.01 \\
 & Part. & 151.4 & \textless\textless0.01 \\
 & Emp. & 146.0 & \textless\textless0.01 \\
\cline{2-4}
 & \%CR & 11.7 & \textless\textless0.01 \\
 & R:Q & 163.5 & \textless\textless0.01 \\
 & MI NonAdh. & 162.8 & \textless\textless0.01 \\
\thickhline
\end{tabular}
\caption{Friedman omnibus tests across the three counsellors within each client framework for the MITI metrics in Table~\ref{tab:miti_results}. All rows use the same matched set of $n=83$ profiles except \%CR under \Name, where $n=70$ because the metric is undefined when no reflections are present.}
\label{tab:miti_omnibus}
\end{table}

%% file: grounding_all_settings_multicolumn.tex
\section{Non-Grounded Rate Across All Settings}
\label{app:grounding_all_settings_multicolumn}
\label{app:grounding_all_settings}

\begin{multicols}{2}
Figure~\ref{fig:secondary_rq} (left) in the main text fixes the counsellor to MIBot v6.3A so that the comparison between Profile-Only and \Name isolates the effect of the client framework. For completeness, this appendix provides the corresponding boxplot across all six experimental arms together with the real-transcript baseline.

\begin{figure}[H]
\centering
\includegraphics[width=0.72\textwidth]{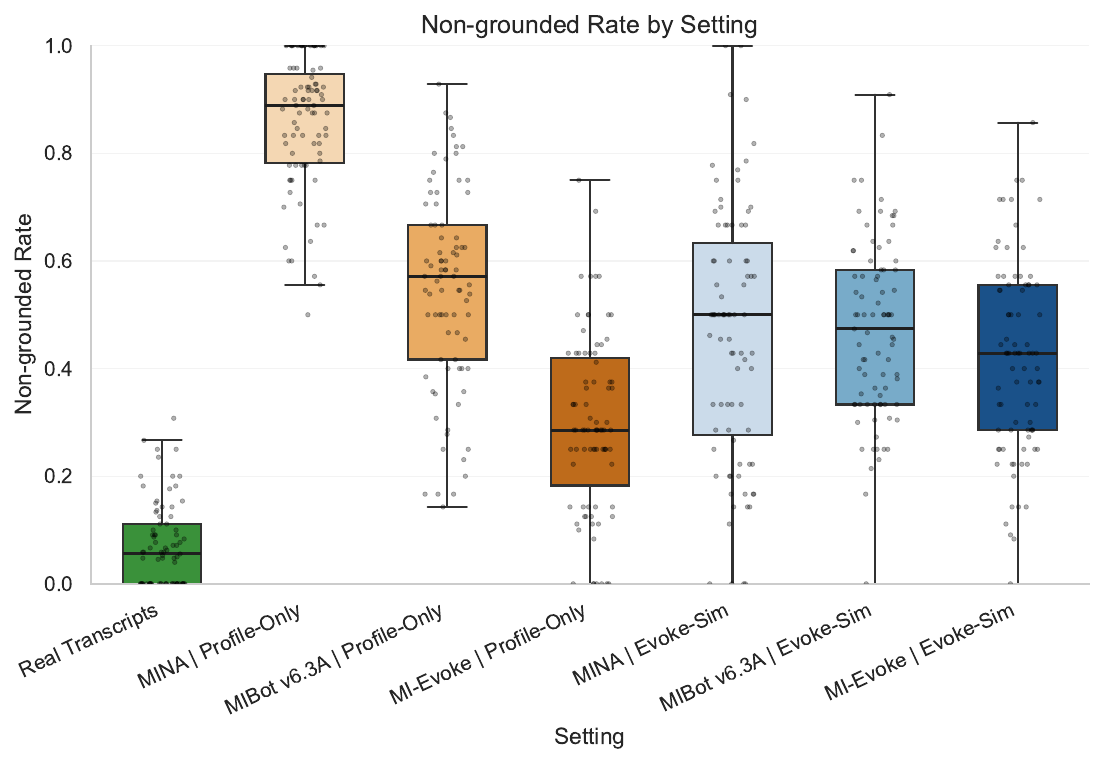}
\caption{Distributions of non-grounded rate across all six experimental arms, together with the real-transcript baseline.}
\label{fig:grounding_boxplot_all_settings}
\end{figure}

This full view is useful for showing how non-grounded rate varies jointly with client framework and counsellor choice. At the same time, it is less targeted for the main RQ2 comparison because counsellor effects and framework effects are mixed together. For that reason, the main text reports the matched-counsellor comparison, while Figure~\ref{fig:grounding_boxplot_all_settings} below serves as the full reference view.
\end{multicols}

%% file: rq2_pairwise_tests_multicolumn.tex
\section{Pairwise Statistical Tests for RQ2}
\label{app:rq2_pairwise_tests_multicolumn}
\label{app:rq2_pairwise_tests}

\subsection{Non-Grounded Rate}
\begin{multicols}{2}
Section~\ref{sec:results} focuses on the matched-counsellor MIBot v6.3A comparison, but Table~\ref{tab:rq2_grounding_pairwise} reports the same Profile-Only vs.\ \Name framework test for all three counsellors. We use paired Wilcoxon signed-rank tests across the same 83 profiles, together with bootstrap 95\% confidence intervals for the mean difference (\Name $-$ Profile-Only).

The pairwise results clarify that \Name lowers non-grounded rate for MINA and MIBot v6.3A, but raises it for \EvokeCounsellor. The matched MIBot v6.3A row is the one directly aligned with the main-text claim in Figure~\ref{fig:secondary_rq}, which is the only comparison for which a human client baseline is also available.
\end{multicols}

\begin{table}[H]
\centering
\small
\setlength{\tabcolsep}{5pt}
\renewcommand{\arraystretch}{1.15}
\begin{tabular}{lccccc}
\thickhline
\textbf{Counsellor} & \textbf{Profile-Only} & \textbf{\Name} & \textbf{Diff.} & \textbf{95\% CI} & \textbf{P-value} \\
\hline
MINA & 0.86 & \textbf{0.46} & -0.40 & [-0.45, -0.34] & \textless\textless0.01 \\
MIBot v6.3A & 0.55 & \textbf{0.47} & -0.079 & [-0.13, -0.03] & \textless\textless0.01 \\
\EvokeCounsellor & \textbf{0.29} & 0.41 & 0.12 & [0.068, 0.17] & \textless\textless0.01 \\
\thickhline
\end{tabular}
\caption{Pairwise comparisons of non-grounded rate between Profile-Only and \Name within counsellor. The same matched set of $n=83$ profiles is used for every row. P-values are Holm-corrected over the three tests in the table. All tests use paired Wilcoxon signed-rank tests. Diff.\ = \Name $-$ Profile-Only.}
\label{tab:rq2_grounding_pairwise}
\end{table}

\subsection{Rate of Reveal Summary Metrics}
\begin{multicols}{2}
To summarize the rate-of-reveal trajectories in Figure~\ref{fig:secondary_rq}, we compare Profile-Only and \Name within counsellor on four per-profile summaries: time to 0.25 revealed fraction, time to 0.50 revealed fraction, time to the last new reveal, and slope until the last reveal. All comparisons use paired Wilcoxon signed-rank tests over the same 83 profiles, with Holm-corrected p-values and bootstrap 95\% confidence intervals for the mean difference (\Name $-$ Profile-Only).

These pairwise tests support the main rate-of-reveal pattern for the two MI-adherent counsellors: under \Name, MIBot v6.3A and \EvokeCounsellor take significantly longer to reach the reveal thresholds, continue revealing new items for longer, and show shallower reveal slopes. MINA shows a different pattern because many \Name conversations terminate early, and its confrontational style leads to many early barrier reveals.
\end{multicols}

\begin{table}[H]
\centering
\small
\setlength{\tabcolsep}{4pt}
\renewcommand{\arraystretch}{1.15}
\begin{tabular}{llccccc}
\thickhline
\textbf{Counsellor} & \textbf{Metric} & \textbf{Profile-Only} & \textbf{\Name} & \textbf{Diff.} & \textbf{95\% CI} & \textbf{P-value} \\
\hline
\multirow{4}{*}{MINA}
& Time to 0.25 & \textbf{5.3} & 3.5 & -1.8 & [-3.7, 0.14] & 0.029 \\
& Time to 0.50 & \textbf{6.1} & 3.5 & -2.5 & [-4.6, -0.48] & 0.01 \\
& Time to Last New & \textbf{6.1} & 3.5 & -2.5 & [-4.6, -0.48] & 0.02 \\
& Slope Until Last & 0.034 & \textbf{0.0012} & -0.033 & [-0.042, -0.023] & \textless\textless0.01 \\
\hline
\multirow{4}{*}{MIBot v6.3A}
& Time to 0.25 & 5.2 & \textbf{16.8} & 11.5 & [9.8, 13.3] & \textless\textless0.01 \\
& Time to 0.50 & 11.8 & \textbf{20.2} & 8.4 & [5.9, 10.9] & \textless\textless0.01 \\
& Time to Last New & 16.1 & \textbf{20.3} & 4.2 & [1.6, 6.7] & 0.014 \\
& Slope Until Last & 0.034 & \textbf{0.015} & -0.019 & [-0.024, -0.014] & \textless\textless0.01 \\
\hline
\multirow{4}{*}{\EvokeCounsellor}
& Time to 0.25 & 5.4 & \textbf{12.9} & 7.4 & [6.5, 8.4] & \textless\textless0.01 \\
& Time to 0.50 & 10.4 & \textbf{13.5} & 3.1 & [2.0, 4.2] & \textless\textless0.01 \\
& Time to Last New & 12.6 & \textbf{13.7} & 1.1 & [-0.12, 2.2] & 0.041 \\
& Slope Until Last & 0.051 & \textbf{0.018} & -0.033 & [-0.038, -0.028] & \textless\textless0.01 \\
\thickhline
\end{tabular}
\caption{Pairwise comparisons of rate-of-reveal summary metrics between Profile-Only and \Name within counsellor. The same matched set of $n=83$ profiles is used for every row. P-values are Holm-corrected over the 12 tests in the table. All tests use paired Wilcoxon signed-rank tests. Diff.\ = \Name $-$ Profile-Only. For the time metrics, positive differences indicate a slower rate of reveal under \Name; for Slope Until Last, negative differences indicate a lower overall rate of reveal under \Name.}
\label{tab:rq2_reveal_pairwise}
\end{table}